\documentclass[twocolumn,onespacing,letterspaper,fleqn,usenatbib]{mnras} 

\usepackage{newtxtext,newtxmath}

\usepackage[T1]{fontenc}
\usepackage{ae,aecompl}

\usepackage[pdftex]{graphicx}	
\usepackage{epstopdf}
\usepackage{amsmath}	
\usepackage{pdflscape}
\usepackage{caption}
\usepackage{threeparttable}
\usepackage{longtable}
\usepackage{threeparttablex}
\usepackage{dcolumn}
\usepackage{multirow}
\usepackage{supertabular, booktabs}
\usepackage{tabularx}
\usepackage{subcaption}
\usepackage{mathtools}
\usepackage{blindtext}
\usepackage{float}    
\restylefloat{table}    
\newcommand{\kms}{km s$^{-1}$}
\newcommand{\ra}{$\alpha$}
\newcommand{\dec}{$\delta$}
\newcommand{\pmra}{$\mu_{\alpha}$}
\newcommand{\pmdec}{$\mu_{\delta}$}

\newcommand{\plx}{$\pi$}
\newcommand{\pkin}{$P_{\rm kin}$}

\newcommand{\ycmd}{{\it Youth(CMD)}}
\newcommand{\ynuv}{{\it Youth(NUV)}}
\newcommand{\lx}{$L_{\rm X}$}
\newcommand{\msun}{$M_{\sun}$}
\newcommand{\bprp}{$G_{\rm BP}-G_{\rm RP}$}
\newcommand{\gbp}{$G_{\rm BP}$}
\newcommand{\grp}{$G_{\rm RP}$}
\newcommand{\mg}{$M_{\rm G}$}
\newcommand{\lxlbol}{log$L_{\rm X}/L_{\rm bol}$}

\title[Low-mass members of NYMGs from {\it Gaia} EDR3]{Low-mass members of nearby young stellar moving groups from {\it Gaia} EDR3}

\author[J. Lee, I. Song and S. J. Murphy]{
Jinhee Lee, $^{1}$\thanks{E-mail: jinhee7493@pusan.ac.kr}
Inseok Song,$^{2}$
Simon J. Murphy$^{3}$
\\
$^{1}$Department of Earth Sciences, Pusan National University, Busan 46241, Republic of Korea\\
$^{2}$Department of Physics and Astronomy, The University of Georgia, Athens, GA 30605, USA\\
$^{3}$School of Science, The University of New South Wales Canberra, ACT 2600, Australia\\
}

\date{Accepted XXX. Received YYY; in original form ZZZ}

\pubyear{2021}

\begin{document}
\label{firstpage}
\pagerange{\pageref{firstpage}--\pageref{lastpage}}
\maketitle

\begin{abstract}
{\it Gaia} EDR3 offers greatly improved kinematics for nearby objects, including members of nearby young stellar moving groups (NYMGs).  In this study, we aim to identify low-mass NYMG members (spectral types of M0 to mid-L) in {\it Gaia} EDR3.  We calculated spatio-kinematic membership probabilities utilising a Bayesian membership probability calculation scheme developed in our previous study.  We evaluated stellar youth primarily based on colour-magnitude diagram positions.  Combining spatio-kinematic membership assessment and youth evaluation, we identified $\sim$2900 low-mass NYMG candidate members including $\sim$700 previously claimed members.  In the set of $\sim$2200 new candidate members, $\sim$550 appear to be young based on NUV brightness.
Our pilot spectroscopic study with WiFeS on the ANU 2.3-m telescope observed 78 candidates, with 79 per cent confirmed as members. Using our new member sample, we estimated an isochronal age of the $\beta$ Pictoris Moving Group. The mean age ($\sim$10 Myr), which is around half the age of recent estimates, suggests either a truly younger age of the $\beta$ Pictoris Moving Group or inaccuracies in contemporary isochrones. As the main results of this study, we provide lists of newly confirmed and candidate low-mass NYMG members.
\end{abstract}




\section{Introduction}

Nearby, young stellar moving groups (NYMGs hereafter) are loose stellar associations whose members are gravitationally unbound \citep{zuc04}.  Their young ages ($\lesssim$150 Myr) and proximity (mean distances $<$ 100 pc) provide unique opportunities in contemporary astronomy.  Members spanning various spectral types with well-defined ages are used in studying empirical isochrones \citep{her15, liu16}.  They have been prime targets for direct imaging of exoplanets and used as a laboratory for studying planetary formation and evolution \citep{mar08, mar10, lag10, cha17}.  
Their brown dwarf members are also excellent analogs to  directly imaged giant exoplanets \citep{fah13a, fah13b}.
Moreover, because their lowest mass members can be identified more easily compared to older and/or more distant stellar groups, NYMGs are also useful for investigating stellar mass functions \citep{kra14, shk17}.  As sparse groups, studying the formation environment of NYMGs is crucial to understanding the evolution and dispersion of clusters \citep{sch12}. \newline

Because of their importance and utility, the identification of NYMGs and their members has been a hot topic over the past two decades (e.g., \citealt{zuc01, zuc04, mur10, sch12, rod13, mal13, gag14, shk17, bel17, gag18a, sch19}).  These co-moving members of NYMGs are identified primarily based on their spatial position and velocity.  While the keys in the identification of NYMG members are the same (i.e., using their spatial position and motion), details in the selection  differ between various authors (e.g., \citealt{sch12, shk12, rod13, mal13, gag14, rie17, cru19}).  One of the most successful NYMG membership identification schemes is the BANYAN series which is based on Bayesian statistics \citep{mal13, gag14, gag18a}.  Similar to the BANYAN method, we have developed a membership probability calculation scheme, the Bayesian Analysis of Moving Groups (BAMG; \citealt{lee18}, Paper I).  BAMG offers several improvements over BANYAN such as using a more realistic field star model (Paper I) and less dependency on the initial choice of members \citep[Paper II]{lee19a}.  The most recent version of the BAMG code updates the groups based on the unsupervised machine learning analysis of nine known NYMGs \citep[Paper III]{lee19b}.  Using 652 previously confirmed members of the nine NYMGs, our unsupervised machine learning analysis recognized eight distinct groups.  In this paper, we utilise the properties of the eight recognized moving groups to identify previously unknown low-mass members included in the {\it Gaia} EDR3 catalogue. \newline

The most recent data release from the {\it Gaia} mission \citep{gai16}, {\it Gaia} Early Data Release 3 (EDR3) \citep{gaiaedr3} enables the identification of very low-mass NYMG members thanks to precise parallax measurements.  The available parallax information can also be used for the assessment of youth ($\lesssim$150 Myr) from colour-magnitude diagrams because young stars and brown dwarfs are over-luminous compared to their older counterparts.  Additionally, {\it Gaia} EDR3 includes RV measurements \citep{sea21} for some stars down to early-M type stars, which makes a complete set of astrometric parameters to calculate robust NYMG membership probabilities.  In this study, we present $\sim$2200 new low-mass (spectral types of M0 to mid-L) NYMG candidate members identified by a combined assessment of their spatio-kinematic membership and youth. \newline

We describe the method in Section 2 followed by the results in Section 3, including the pilot spectroscopic observations.  The population statistics of young nearby objects are discussed in Section 4, including a reassessment of the isochronal age of the $\beta$ Pictoris Moving Group using the newly confirmed membership. The conclusion and summary of this paper are presented in Section 5.

\section{Method}

To be assessed as a new member of a NYMG, a candidate's position, velocity, and age should be consistent with those of known NYMG members.  For identifying low-mass NYMG members, we apply two restrictions to {\it Gaia} EDR3 at the beginning of this study: distance and colour.  We limit distance to $\leq$150 pc and \bprp\ colour to $\geq$1.8, which corresponds to a spectral type of M0V or later.  These two restrictions retain $\sim$1.5 million sources.  Beginning with this initial dataset, candidates are selected based on their spatio-kinematic membership and a youth assessment. 

\subsection{Spatio-kinematic membership assessment}

Spatio-kinematic membership is assessed using the Bayesian Analysis of Moving Groups code (BAMG\footnote{Calculation is available at http://movinggroups.org}; \citealt{lee18, lee19a}, Paper I and II).  This tool requires ellipsoidal models of the NYMGs under consideration. 
 The ellipsoidal models in positional space ($XYZ$) and velocity space ($UVW$) are treated independently, which assumes the correlation between spatial and kinematic properties is small enough to be neglected.
 A recent study by \citet[Paper III]{lee19b} evaluated all known NYMGs ($<$150 Myr and a mean distance $<$100 pc: 9 groups) and identified 8 groups: TW Hydrae Association (TWA), $\beta$ Pictoris Moving Group (BPMG),  Carina Association (Carina), Argus Association (Argus), AB Doradus Moving Group (ABDor), and Volans-Carina Association (VCA), 32 Ori-Columba Association (ThOrCol), Tucana-Horologium-Columba Association (TucCol).  Using an iterative model construction method described in Paper II, we constructed ellipsoidal models of these 8 NYMGs.  Table~\ref{tab:models} describes the parameters of our adopted ellipsoidal NYMG models.

\begin{figure}
\includegraphics[width=0.99\linewidth]{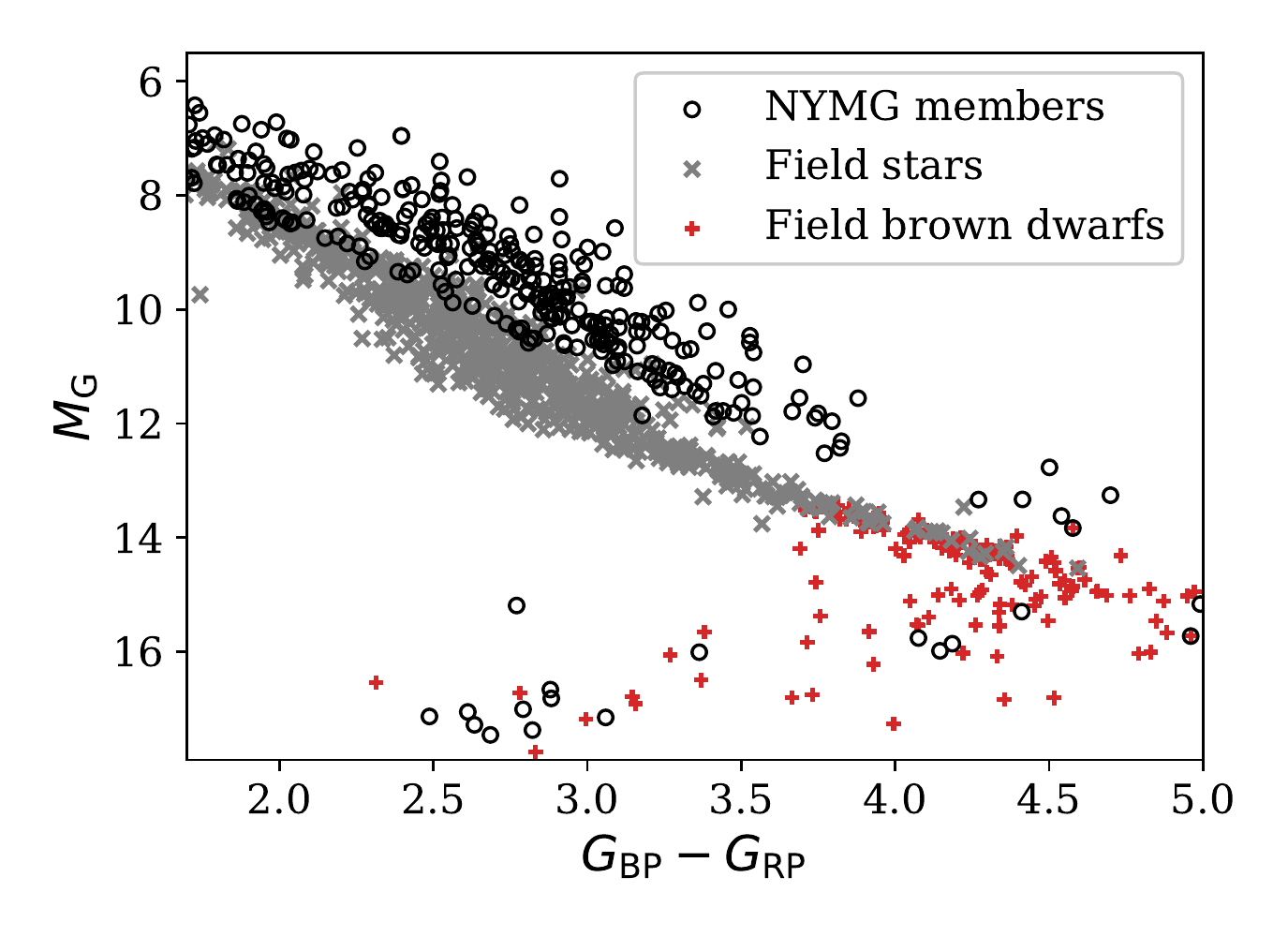}
\caption{Distribution of field stars and NYMG members in Gaia $G$, \gbp\ and \grp\ bandpasses. NYMG members were taken from Paper II \citep{lee19a}, and the field stars were taken after cleaning {\it Gaia} EDR3 utilising criteria suggested by \citet{lin18}.
Field brown dwarfs were taken adopting the procedure by \citet{rey18}.}
\label{fig:ymap1}
\end{figure}

\begin{figure}
\includegraphics[width=0.99\linewidth]{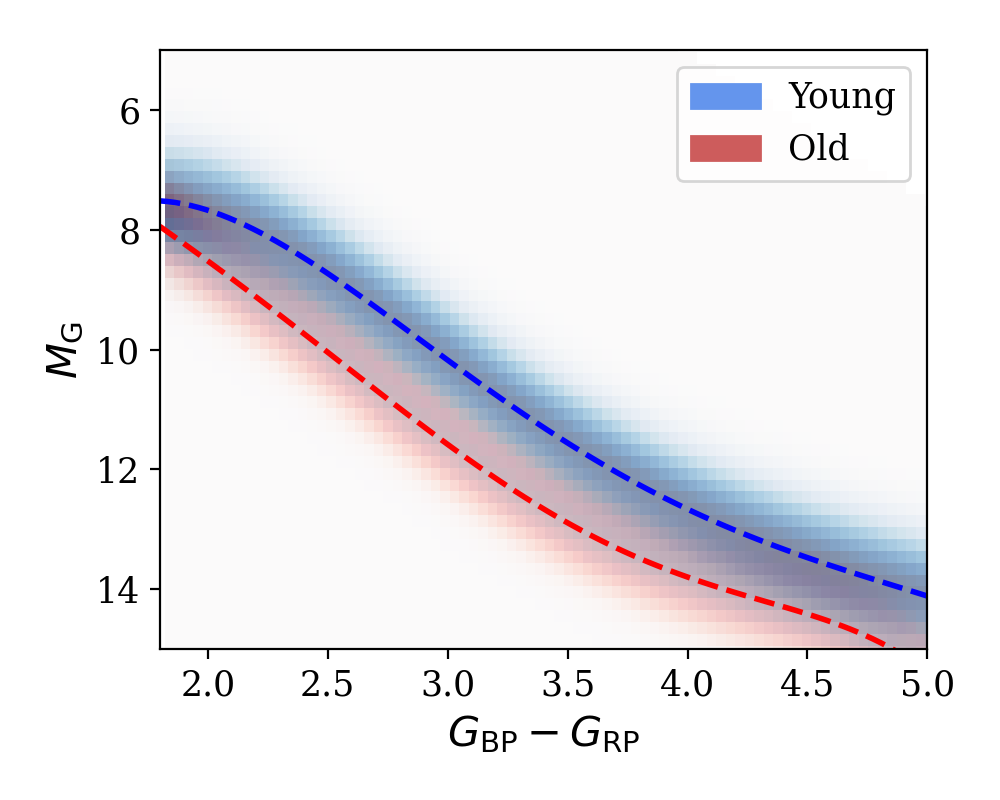}
\caption{A model map for evaluating youth from the CMD.  Details on the production of this map are explained in Appendix A. }
\label{fig:ymap4}
\end{figure}

\begin{table*}
\centering
\small
\setlength\tabcolsep{3pt} 
\begin{threeparttable}
\begin{tabular}{crrrrrrrrrrrrrrrrrr}
\toprule
Name & $X$ & $Y$ & $Z$ & $\sigma_{\rm X}$ & $\sigma_{\rm Y}$ & $\sigma_{\rm Z}$ & $\phi_{\rm xyz}$ & $\theta_{\rm xyz}$ & $\psi_{\rm xyz}$ & $U$ & $V$ & $W$ & $\sigma_{\rm U}$ & $\sigma_{\rm V}$ & $\sigma_{\rm W}$ & $\phi_{\rm uvw}$ & $\theta_{\rm uvw}$ & $\psi_{\rm uvw}$  \\
\cmidrule(lr){2-4}\cmidrule(lr){5-7}\cmidrule(lr){8-10}\cmidrule(lr){11-13}\cmidrule(lr){14-16}\cmidrule(lr){17-19} 
& \multicolumn{3}{c}{pc} & \multicolumn{3}{c}{pc} & \multicolumn{3}{c}{$\circ$} & \multicolumn{3}{c}{\kms} & \multicolumn{3}{c}{\kms} & \multicolumn{3}{c}{$\circ$} \\
\hline 
      TWA   &  15.0  & -48.9 &  23.7 &  12.6 &   6.3 &   5.2 & -134 &  -14 &   7 & -12.4 & -18.6 &  -5.8 &   2.7 &   1.6 &   1.1 & -134 &  -11 &  -54   \\ 
      BPMG & 25.4 &  -0.2  & -20.7 &  31.2 &  15.8 &  8.0 &  -166 &  -11 &  -21 & -9.4 & -15.8 &  -8.8 &   2.5 &   1.7 &   1.0 & -180 &   24 &  -57   \\ 
   ThOrCol & -47.5 & -41.6 & -28.8 &  45.2 &  16.0 &  6.0 & -161 &    3 &   21 & -11.2 & -21.7 &  -7.9 &   4.8 &   2.7 &   2.4 & -134 &  46 &  69 \\ 
    TucCol  &  17.3  & -24.0 & -38.2 &  19.1 &  10.5 &   3.3 &  124 &  -36 &   20 &  -9.1 & -20.6 &  -2.7 &   4.4 &   2.6 &   1.7 & -172 &   44 &  -47  \\ 
    Carina  &  -6.0   & -86.0 & -36.4 &  51.0 &  26.8 &  20.5 & -129 & -40 &  -158& -12.4 & -21.7 & -4.1 &   3.6 &   2.9 &   1.8 & 175 &  13  & -34   \\ 
     Argus  &  10.7 & -121.3 & -19.9 &  37.2 &  23.6 &  6.1 &   116 &  -4  &   0    & -23.0 & -14.0 &  -5.9 &   2.8 &   2.2 &   1.2 &  -156 &  7   &  25   \\ 
     ABDor &  -0.9   &   3.5 & -11.0 &  28.3 &  23.7  &  22.8 & -163 &  18  & 76   &   -12.1 & -23.4 & -9.7 &  10.6 &  4.8 &   3.9 & -149 & 27 &  11   \\ 
       VCA  &  20.9  & -83.2 & -15.7 &   4.1 &   3.0 &   1.9  &  -72  &   8   &   -72  & -16.0 & -29.6 &  -1.1 &   5.4  &   0.9 &   0.6 & -103 & 8   &  15  \\ 
     \bottomrule
\end{tabular}
\end{threeparttable}
\caption{The ellipsoidal model parameters for the eight NYMGs considered in this study.  $X, Y, Z, U, V$, and $W$ are centre positions of the models.  $\sigma$ values are the standard deviation of the models, and angles represent Euler angles.}
\label{tab:models}
\end{table*}

In the spatio-kinematic membership probability (\pkin) calculation, a candidate's six astrometric parameters are required: R.A., Dec., proper motions in R.A. and Dec., distance, and radial velocity (RV).  In the case of missing RV, a less robust membership probability can still be calculated by marginalising over RV.  In {\it Gaia} EDR3, only a subset of sources has RV measurements ($\sim$31,000 of the $\sim$1.5 million sources).

To be considered as potential candidate members, we adopt a threshold of \pkin$\geq$90 per cent as was used in Paper II.  If an EDR source has \pkin$\geq$90 per cent in any of the eight NYMGs considered in this paper, then the source is retained as a candidate member.  Sources having a marginal \pkin\ value ($<$90 per cent) for a group but the sum of \pkin\ across multiple groups over the threshold ($\Sigma$\pkin$\geq$90) are retained; they are considered as having ambiguous membership status for a single group. For more details on the BAMG algorithm and the definition of these moving groups, we refer the reader to Papers I, II, and III and references therein \citep{lee18, lee19a, lee19b}.

\subsection{Youth assessment}

In 6-D position and velocity space, NYMG members show overdensities in certain regions. However, even in such ranges, old field stars appear to be the dominant population. This means that any membership survey solely based on spatio-kinematics (i.e., position and velocity) will suffer from false positives from old interlopers. Therefore, an assessment of youth is a pivotal step in any reliable NYMG membership survey. Young low-mass stars have strong surface magnetic fields manifested as strong X-ray emission and/or chromospheric UV emission \citep{kas97, rod13}.  Strong H$\alpha$ emission and stronger Li absorption  compared to field stars are also used as signs of youth \citep{zuc04, son12, mur13}.  In the low-mass regime, young objects ($\lesssim$150 Myr) are still contracting after their formation \citep{bur01, mar07},  making them over-luminous on colour-magnitude diagrams (CMDs).  The high-quality parallaxes and sensitive photometry of {\it Gaia} EDR3 can therefore be used to readily select young object candidates.

We evaluated youth ($\lesssim$150 Myr) primarily based on CMDs with $NUV$ magnitudes from {\it GALEX} DR5 \citep{bia11} and X-ray luminosities (\lx) from the Second ROSAT All-Sky Survey Point Source Catalog (2RXS; \citealt{bol16}) used as supplementary evidence of youth. 
A CMD of bona fide NYMG members from Paper II and field stars is shown in Fig.~\ref{fig:ymap1}.  The NYMG members are well separated from field stars in the range \mg\ $\leq$14 mag. At \mg$\gtrsim$14 mag, \bprp\ colours of known NYMG members which are mostly brown dwarfs become bluer. 
 \mg=14 corresponds to 2500-2700 K (BHAC15; \citealt{bar15}). The objects with an effective temperature cooler than 2700 K are defined as ultracool dwarfs \citep{kir97}, which correspond to spectral types of M7 or later.
In this study, we split candidate members at \mg=14 for effective youth assessment [\mg$\leq$14 mag; cool dwarfs (M0 to mid-M), \mg$>$14 mag; ultracool dwarfs (late-M to mid-L)].

\subsubsection{Cool dwarfs}

Stellar youth ($\lesssim$150 Myr) for cool dwarfs was evaluated primarily based on the \bprp\ vs \mg\ colour-magnitude diagram (Fig.~\ref{fig:ymap1}), which well separates NYMG members (10-150 Myr) from old field stars. However, these young and old stars cannot be perfectly distinguished at a sharp boundary, rather there is a combination of some young and old stars at the boundary region; this combination might be caused by uncertainties on data (colour, magnitude, and parallax), multiplicity, and/or a mixture of young field stars.  

Utilising this characteristic on the CMD, we created a parameter, \ycmd\ to quantify how close a star's position on the CMD is to that of young stars. The \ycmd\ value does not reflect the true likelihood of the young and old star populations. The value is taken from a model map (Fig.~\ref{fig:ymap4}) representing the ratio of young and field stars assuming an equal population of young and field stars at a given colour on the CMD. The \ycmd\ value ranges from 0.0 to 1.0, which corresponds to the CMD position of old and young stars, respectively. Young stars have a \ycmd\ value close to 1.0.  Details on the production of the \ycmd\ map are given in Appendix A.

Unresolved old binaries or multiple systems in EDR3 can be misidentified as potential young stars from the \ycmd\ analysis. The angular resolution of the {\it Gaia} observations implies that binaries with $\gtrsim$0.4 arcsec separation would have been resolved as two sources. 
Unresolved, close binaries (separation$<$0.4 arcsec) can have excess astrometric noise which represents the disagreement between the observations and the best-fitting standard (e.g., single) astrometric model. {\it Gaia} catalogue parameters ({\it astrometric\_excess\_noise} and {\it astrometric\_excess\_noise\_sig}), or the ratio of these two parameters \citep{lin21}, can potentially be used to flag unresolved binaries among EDR3 sources. Assuming {\it Gaia} EDR3 sources with measured RVs without noticeable variation in RV are single stars, we compared the ratio of {\it astrometric\_excess\_noise} and {\it astrometric\_excess\_noise\_sig} among NYMG candidates with and without measured RVs. We found no meaningful difference in the comparison.  
In conjunction with the fact that young and old sequences on the CMD (dashed lines in Fig.~\ref{fig:ymap4}) are well separated (1$-$1.5 magnitudes), this result likely implies that the contamination level in our selected NYMG candidates due to old, unresolved binaries is small. \newline

\begin{figure*}
\includegraphics[width=0.95\linewidth]{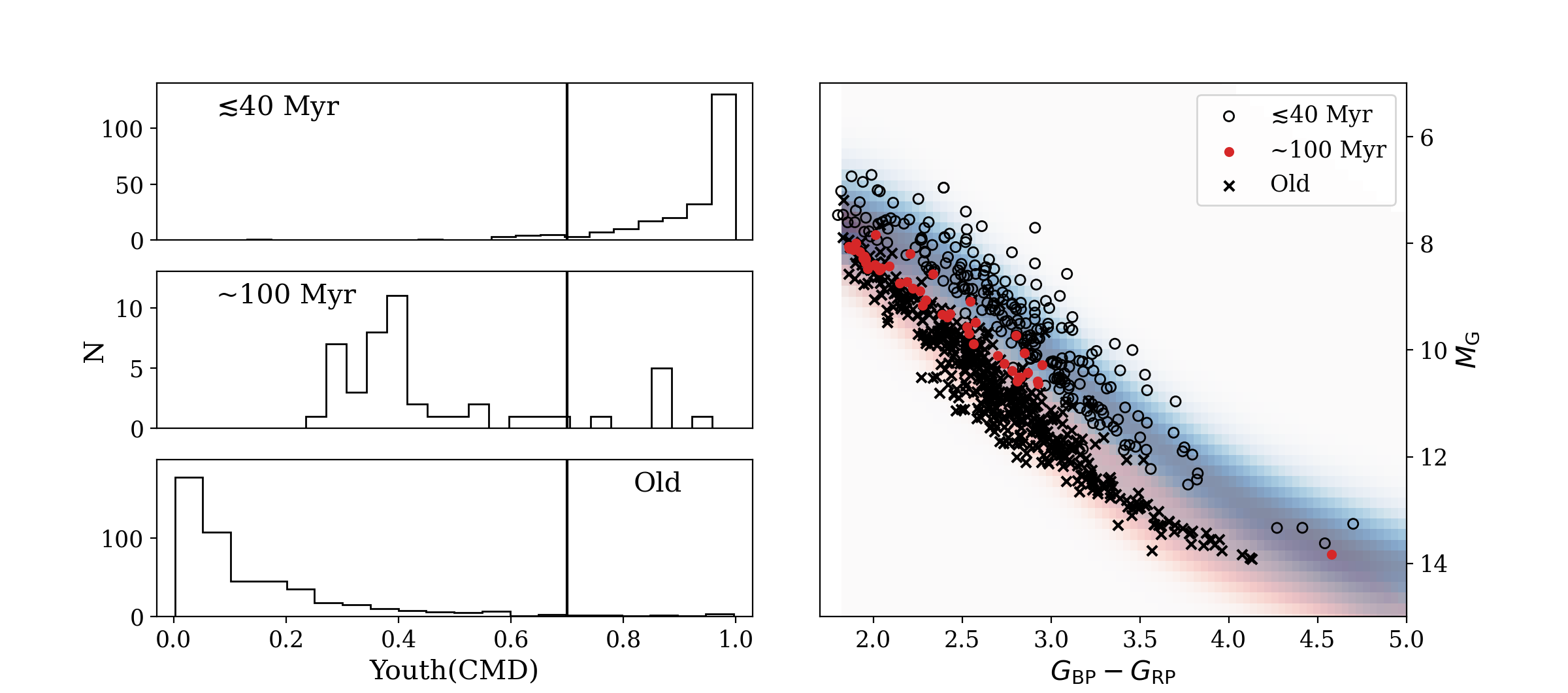}
\caption{Left: histograms of \ycmd\ for 3 age groups: (1) stars belonging to groups younger than 40 Myr, (2) stars belonging to groups ABDor or VCA ($\sim$100 Myr), and (3) randomly selected field stars (old). The vertical line indicates the selection threshold of young candidates [\ycmd=0.7]. Right: a CMD of stars in these 3 age groups.  The background illustrates the model map for obtaining \ycmd\ identical to Fig.~\ref{fig:ymap4}.}
\label{fig:pyouth_nymg}
\end{figure*}

To use the \ycmd\ parameter in our NYMG candidate member selection procedure, the determination of a proper threshold \ycmd\ for evaluating youth is required.  To do this we took bona fide NYMG members and 500 randomly selected stars from {\it Gaia} EDR3 as {\it young} and {\it old} samples, respectively.  We then split the young star sample into two subsets: (1) $\lesssim$40 Myr-old groups (TWA, BPMG, ThOrCol, TucCol, Carina, and Argus) and (2) $\sim$100 Myr-old groups (ABDor and VCA).  The stars of $\sim$100 Myr-old groups are largely mixed with field stars on the CMD (see the right panel of Fig.~\ref{fig:pyouth_nymg}), and we expect a large level of contamination when candidate members are selected  solely based on the CMD.  Using these samples we explored which threshold \ycmd\ value should be used for an effective selection of young candidates (i.e., reducing contamination of field stars).
Fig.~\ref{fig:pyouth_nymg} presents \ycmd\ values for these 3 age groups ($\lesssim$40 Myr, $\sim$100 Myr, and old).  The samples with ages $\leq$40 Myr have \ycmd\ values larger than 0.70 in most cases.  The major population of the 100 Myr-old samples have \ycmd\ values 0.25$-$0.45.  Most of the old samples have \ycmd\ smaller than 0.20.

As can be seen in the left panel of Fig.~\ref{fig:pyouth_nymg}, selection with \ycmd\ $\geq$0.70 would allow a low contamination rate while \ycmd\ $\geq$0.25  would contain a significant number of old interlopers. In this study, we adopt a more stringent \ycmd\ threshold of 0.70 to limit the contamination by old interlopers.  However, we note that it would probably miss many candidate members for VCA and ABDor groups, for example.

\begin{figure*}
\begin{subfigure}[b]{.49\linewidth}
\includegraphics[width=0.95\linewidth]{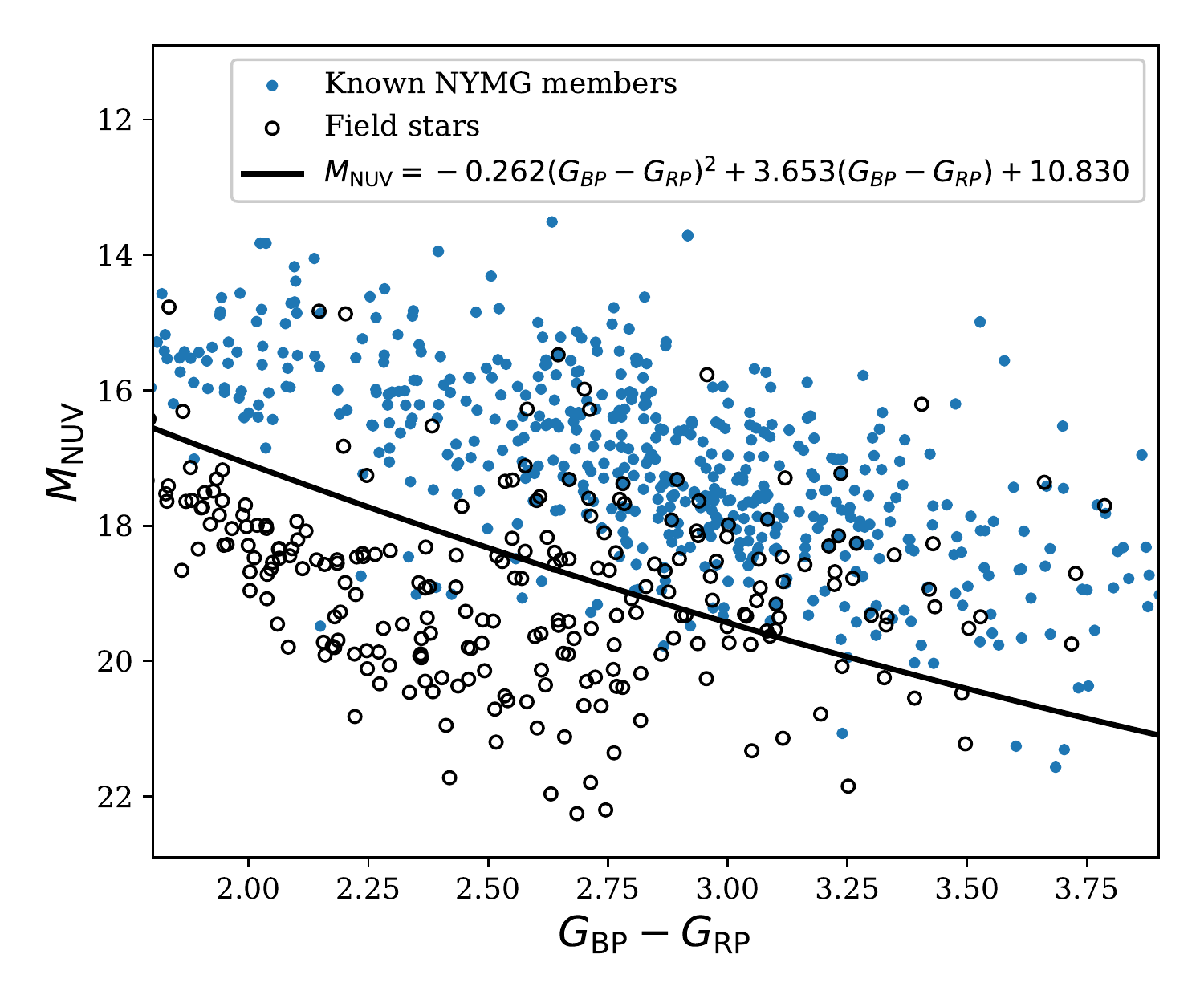}
\end{subfigure}
\begin{subfigure}[b]{.49\linewidth}
\includegraphics[width=0.95\linewidth]{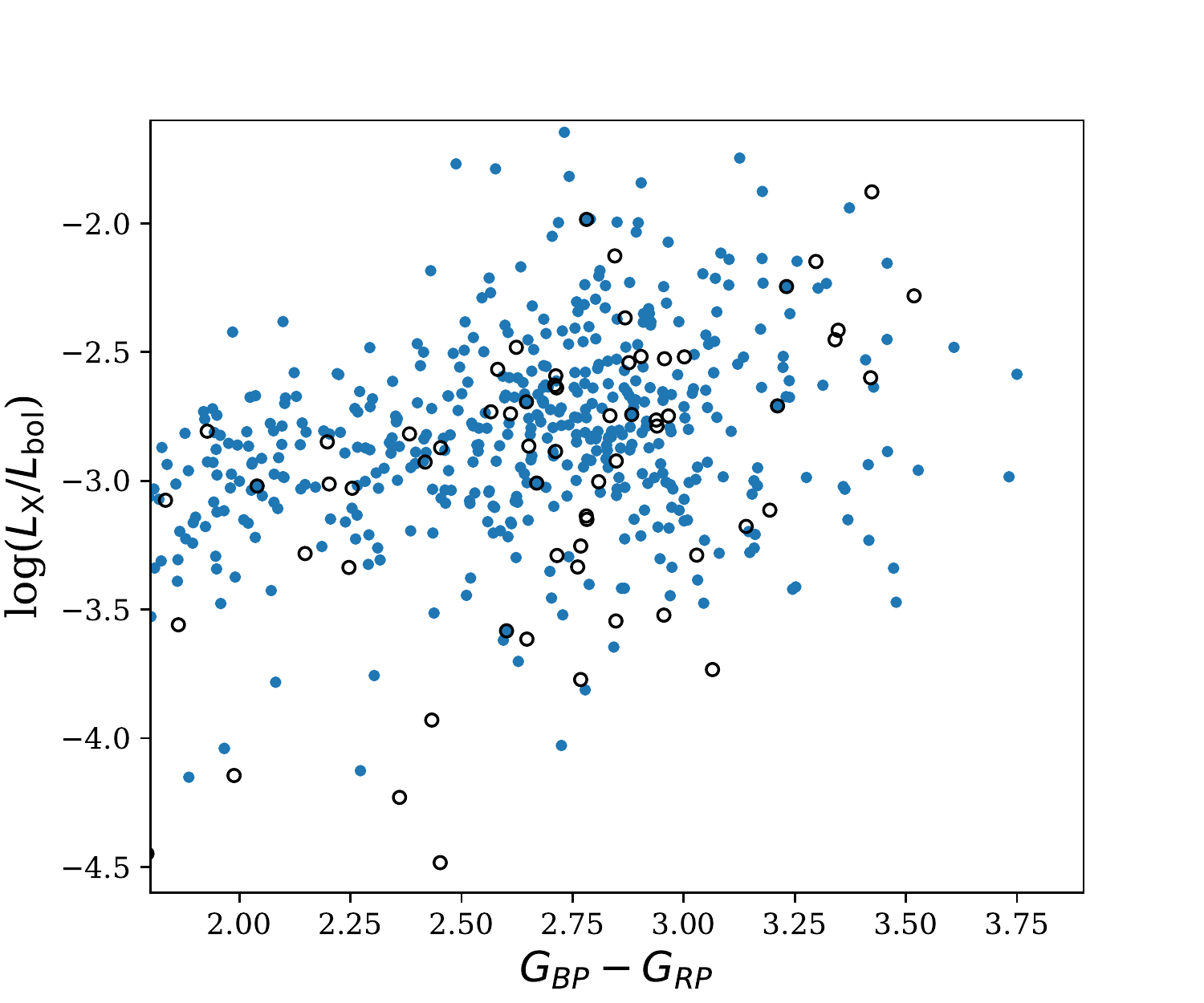}
\end{subfigure}
\caption{$NUV$ magnitudes (left) and the fraction of X-ray luminosities \lxlbol\ (right) of known NYMG members (blue circles) and field stars (black empty circles). Left: we established an empirical boundary based on a visual inspection that separates young stars from old stars. Stars brighter than the boundary are evaluated as young [\ynuv=1] while stars fainter than the line are old [\ynuv=0]. 
Right: young and field stars are not distinguishable, hence \lxlbol\ is not used in youth evaluation. Details are explained in Section 2.2.1.}
\label{fig:nuvlx}
\end{figure*}

As supplementary evidence for youth evaluation, we considered $NUV$ magnitude and the fraction of X-ray luminosity \lxlbol. In Fig.~\ref{fig:nuvlx} (left), young and old stars are distinguished around an empirically determined boundary line. 
Stars brighter than the boundary are young [\ynuv=1], while those fainter than the line are likely old [\ynuv=0]. Stars that are either faint in $NUV$ or near the Galactic plane ($|b|<$ 15$^{\circ}$) are not in the {\it GALEX} DR5 catalog. We considered an upper limit for them taking the sensitivity limit of {\it GALEX} ($NUV$=25.3 mag\footnote{http://www.galex.caltech.edu/researcher/faq.html}). Stars with an upper limit fainter than the boundary are considered as old [\ynuv=0], otherwise, their youth cannot be evaluated [\ynuv=2]. 
On the other hand, \lxlbol\ appears not to be useful distinguishing young stars from old ones in this low-mass regime. 
While \lxlbol\ or the fraction of the X-ray luminosity \lxlbol\ can be an effective age indicator for stars in the spectral type ranges from F to M0 \citep{sch13}, lower mass stars remain saturated in X-ray emission (\lxlbol$\sim-$3.0) for very long time and X-ray emissions cannot be used as an effective age indicator as shown in the right panel of Fig.~\ref{fig:nuvlx}.

\subsubsection{Ultracool dwarfs (UCDs)}

Because of the sparse distribution of known young ultracool dwarfs, we cannot establish the same \ycmd\ map in the ultracool dwarf regime of the CMD (\mg$>$14 mag).
To develop a selection scheme for NYMG ultracool dwarfs, we first eliminated sources that have large astrometric or photometric errors. These high-sigma astrometric-photometric outliers were selected following the procedure in \citet{gaiaedr3}.
They present new ultracool dwarf candidates adopting the selection procedure by \citet{rey18}.  Sources with large flux errors in the $G$ and \grp\ bands and large astrometric errors were classified as outliers.

\begin{figure*}
\includegraphics[width=0.95\linewidth]{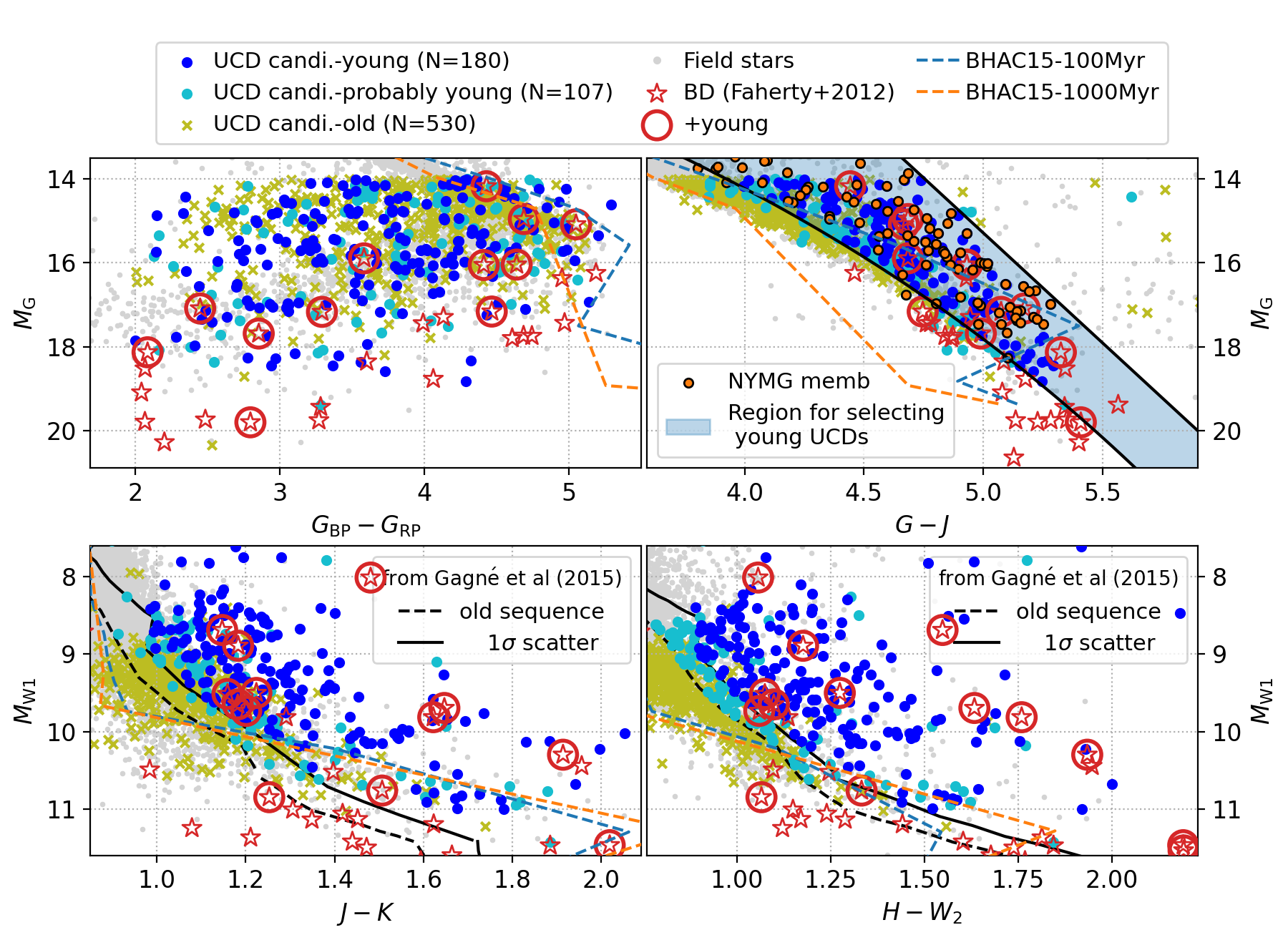}
\caption{CMDs for evaluating the youth of ultracool dwarfs (UCDs).
Field stars are illustrated as grey dots while brown dwarfs in \citet{fah12} are marked as red stars and circled for those young objects.
Among the spatio-kinematic candidates (\pkin$\geq$90 per cent), young, probably young, and old candidates are marked as blue dots, cyan dots, and olive crosses, respectively.
The lines for youth evaluation and isochrones (BHAC 15; \citealt{bar15}) are illustrated.
See text for details on youth evaluation using these diagrams.}
\label{fig:bd_candidates}
\end{figure*}

After filtering these outliers, we evaluated the youth of these spatio-kinematic candidate members using three colour-magnitude diagrams. As mentioned in \citet{gaiaedr3}, \gbp\ is unreliable but the $G-J$ colour is useful in the ultracool dwarf regime. Using the known faintest NYMG members and young brown dwarfs from \citet{fah12}, we defined empirical sequences separating young objects from old ones (upper right in Fig.~\ref{fig:bd_candidates}). The empirical boundaries are presented in the figure and equations ~\ref{eq:bd_young_bottom} and ~\ref{eq:bd_young_top}. 

\begin{equation}
\!\begin{multlined}[t][.75\columnwidth]
  M_{\rm G} = 0.10062248 (G-J)^3 - 0.72214455  (G-J)^2 \\+ 3.95480165 (G-J) + 3.52007669
  \label{eq:bd_young_bottom}
  \end{multlined}
\end{equation}

\begin{equation}
M_{\rm G} = 5.25149246 (G-J) - 10.97509338
  \label{eq:bd_young_top}
\end{equation}
 
Objects located between the boundaries (the blue shaded region in Figure~\ref{fig:bd_candidates}) are considered as young candidates.

In the ultracool dwarf regime, broadband infrared colours have been useful in the selection of young objects \citep{fah12, fah16, gag14, gag15, sch17}. \citet{gag15} provides the old sequence and its 1$\sigma$ scatter on $J-K$ vs $M_{W1}$ and $H-W_2$ vs $M_{W1}$ diagrams (black dashed and solid lines in lower panels of Fig.~\ref{fig:bd_candidates}). 
Objects brighter than the 1$\sigma$ scatter line are considered as young objects while those fainter than the old sequence line are considered as old ones.
The youth assessment criteria incorporating the use of the three CMDs are as follows:

\begin{itemize}
\item Young: evaluated as young objects in all three CMDs.
\item Probably young: evaluated as young objects in two CMDs, but not evaluated as old in any CMD.
\end{itemize}

Similar to the cool dwarfs, the absolute $NUV$ magnitude was used as supplementary evidence for evaluating youth. Note that ultracool dwarfs' intrinsic faintness can limit severely detection in {\it GALEX}.

\section{Results}

Following the procedures described in Section 2, we identified new NYMG candidate members based on the spatio-kinematic candidate selection and youth assessment.

\subsection{Identification of new NYMG members}

\begin{table}[h]
\centering
\small
\begin{threeparttable}
\begin{tabular}{cc}
\toprule
\multicolumn{2}{c}{NYMG members from literature=1,921}  \\
\hline
\multicolumn{2}{c}{\bprp$\geq$1.8 $\cap$ distance $\leq$150 pc=1,115} \\
\hline
Cool dwarfs = 981 & Ultracool dwarfs = 134 \\
\cmidrule(lr){1-1}\cmidrule(lr){2-2}
 \pkin$\geq$90 = 836  &  \pkin$\geq$90 = 109\\
 \ycmd$\geq$0.7 = 651 & young = 69 \\
 \cmidrule(lr){1-1}\cmidrule(lr){2-2}
$\cap$=570 & $\cap$=56 \\
Recovery rate=0.58 & Recovery rate=0.42 \\
\hline
\end{tabular}
\end{threeparttable}
\caption{The numbers of NYMG members from the literature that pass the selection criteria of this study.}
\label{tab:recovery}
\end{table}

\begin{table}[H]
\centering
\small
\begin{threeparttable}
\setlength\tabcolsep{3pt} 
\begin{tabular}{lcc}
\toprule
Criteria & Cool dwarfs = 570 & Ultracool dwarfs = 56 \\
\cmidrule(lr){1-1}\cmidrule(lr){2-2}\cmidrule(lr){3-3}
\ynuv=1 (young)                & 342   & 6 \\ 
\ynuv=0 (old)                    &  141     & 39 \\
\ynuv=2 (indecisive)         &  87     & 11 \\
\hline
\end{tabular}
\end{threeparttable}
\caption{The numbers of recovered NYMG members from the literature that satisfy the supplementary $NUV$ youth criterion.}
\label{tab:known_nuvlx}
\end{table}

The number of EDR3 sources in the initial dataset after the two restrictions (distance $\leq$150 pc and \bprp\ $\geq$1.8) is $\sim$1.5 million.  These sources were split into two types of objects: cool dwarfs and ultracool dwarfs at \mg=14 mag.  The numbers of retained cool dwarfs and ultracool dwarfs are $\sim$0.7 million and $\sim$0.8 million, respectively.

NYMG candidate members were identified and then cross-matched with previously known NYMG members.
We collected previously known NYMG members from $\sim$60 source papers, including the group discovery papers (e.g., TWA; \citealt{kas97}, BPMG; \citealt{zuc01}), surveys searching for low-mass members (e.g., \citealt{gag15, ell16, shk17, gag18b}), and recent papers (e.g., \citealt{sch19, rie19}).
We considered proposed members in the literature if they were suggested not only based on kinematic parameters but also on clear signs of youth. The number of these claimed members is $\sim$2,000, and 1,115 pass our distance and color criteria for the initial data set (\bprp$\ge$1.8 and distance $\leq$150 pc).  
Table~\ref{tab:recovery} presents the number of previously claimed NYMG members that pass the selection criteria of this study.

Among cool dwarfs ($N$=981), 85 per cent ($N$=836) were recovered by \pkin\ selection while 66 per cent ($N$=651) passed the \ycmd\ selection, yielding a final recovery rate of 58 per cent ($N$=570). 
When the supplementary youth criterion using $NUV$ was applied, 342 stars were evaluated as young (Table~\ref{tab:known_nuvlx}).

Amongst the set of 134 known ultracool dwarf NYMG members, 81 ($N$=109) and 51 ($N$=69) per cent passed our \pkin\ selection and the primary youth assessment, respectively.  Fifty six members (42 per cent) were retained by both criteria. 
 Due to their intrinsic faintness in $NUV$, only 7 ultracool dwarf candidates were detected in the NUV band of {\it GALEX}. Therefore, we did not consider NUV data in the youth evaluation of ultracool dwarf candidates. For those seven candidates with NUV detections, we utilized the NUV data in the same manner as in Section 2.2.1 (Table~\ref{tab:known_nuvlx}).

Considering claimed NYMG members in the literature, only 56 per cent (626 out of 1,115) were recovered in our analysis (results from \pkin\ and \ycmd\ analysis, not including the $NUV$ information). Different spatio-kinematic models of NYMGs may be the reason for missing 15 per cent of these literature members, while 35 per cent of them were not selected because of our more systematic, selective youth evaluation method.

\subsubsection{Cool dwarf members}

\paragraph*{Spatio-kinematic assessment} \mbox{}\\
Of the approximately 700,000 cool dwarfs which pass our initial colour and distance cuts, about ten thousand ($N$=10,800) have spatio-kinematic membership probabilities (\pkin) larger than 90 per cent.  A subset of 1,018 stars has RV values.

\paragraph*{Youth assessment}  \mbox{}\\
About forty thousand ($N$=41,636) cool dwarfs which pass our initial colour and distance cuts have \ycmd\ larger than 0.7.  A subset of 3,217 stars has RV values.

\paragraph*{NYMG candidate selection combining \pkin\ and \ycmd}  \mbox{}\\
The spatio-kinematic NYMG candidates and young candidates were cross-matched, retaining 2,611 stars, including 203 stars with RV measurements. These 203 stars have full astrometric parameters and confirmed youth based on a CMD, and we call them highly likely members.  However, because of potential binary contamination, these stars still need to be considered as candidate members until confirmed by follow-up spectroscopic observations.

Among these 2,611 candidate members, 609 have been claimed in the literature. The remaining 2,002 stars are therefore new NYMG candidate members, and details of the entire sample are provided in online tables. The 139 new highly likely members having RVs are listed in Table~\ref{tab:newbonafide1} while the remaining 1,863 new candidates without RVs are in  Table B1.

\begin{table*}
\centering
\begin{threeparttable}
\scriptsize
\setlength\tabcolsep{3pt} 
\begin{tabular}{*{15}{c}}
\hline
{\it Gaia} EDR3 source\_id &Group & \ra\ & \dec\  &  \pmra\ & \pmdec\ & \plx  & RV & $G$ & \bprp & $\sum{P_{\rm kin}}$& \pkin & \ycmd & $NUV$ & \ynuv \tnote{a} \\
& & hh:mm:ss & dd:mm:ss & mas yr$^{-1}$  & mas yr$^{-1}$ & mas  &   \kms & mag & mag & \% & \% &  & mag \\ \hline    
 2364722818954076288 &      ABDor &  00:10:43.18 &  -20:39:08.1 &  120.47 &  -76.52 &   25.41 &     1.3 & 12.20 &  2.53 &  99 &  51\tnote{b} & 0.74 & 19.80 & 1  \\ 
 4980868526585951232 &      ABDor &  00:42:11.04 &  -42:52:55.2 &   83.27 &  -42.64 &   18.92 &     6.6 & 12.30 &  2.42 &  98 &  98 & 0.89 & 19.85 & 1  \\ 
 2575842627879030656 &      ABDor &  01:51:20.06 &  +13:24:49.5 &   73.35 & -173.98 &   26.38 &    -6.0 & 11.07 &  2.24 &  99 &  99 & 0.93 & 18.85 & 1  \\ 
 5179399816430035456 &      ABDor &  03:03:21.47 &  -08:05:16.0 &  119.76 &  -37.69 &   23.95 &    31.7 & 11.83 &  2.50 &  93 &  93 & 0.96 &   -- & 0  \\ 
  501985336493825792 &      ABDor &  04:32:57.54 &  +74:06:58.0 &   78.44 & -124.53 &   29.42 &    -7.3 & 11.07 &  2.27 &  99 &  99 & 0.72 & 18.69 & 1  \\ 
\multicolumn{15}{c}{\dots} \\
\multicolumn{15}{c}{\dots} \\
\multicolumn{15}{c}{\dots} \\
\hline
\end{tabular}
\begin{tablenotes}
\item[a] 0: old, 1: young, 2: indecisive of youth; the youth was evaluated using $M_{\rm NUV}$ (see Section 2.2.1 for details).
\item[b] While the star's \pkin\ of a single group is less than 90 per cent, the combined \pkin\  of groups ($\Sigma$\pkin) is larger than 90 per cent.  The star has an ambiguous status being a member of a single group.
\end{tablenotes}
\end{threeparttable}
\caption{Newly suggested highly likely members of NYMGs identified in this study. All have full kinematic parameters, including EDR3 RVs. The entire list is available online.}
\label{tab:newbonafide1}
\end{table*}

\begin{table}
\centering
\small
\begin{threeparttable}
\setlength\tabcolsep{3pt} 
\begin{tabular}{lcc}
\toprule
Criteria & Cool dwarfs=2002 & Ultracool dwarfs=231 \\
\cmidrule(lr){1-1}\cmidrule(lr){2-2}\cmidrule(lr){3-3}
\ynuv=1 (young)                & 522     & 33 \\ 
\ynuv=0 (old)                    &  408     & 58    \\
\ynuv=2 (indecisive)         &  1072   & 140 \\
\hline
\end{tabular}
\end{threeparttable}
\caption{The numbers of new candidate NYMG members from this study that satisfy the supplementary $NUV$ youth criterion.}
\label{tab:new_nuvlx}
\end{table}

\begin{figure*}
\includegraphics[width=0.9\linewidth]{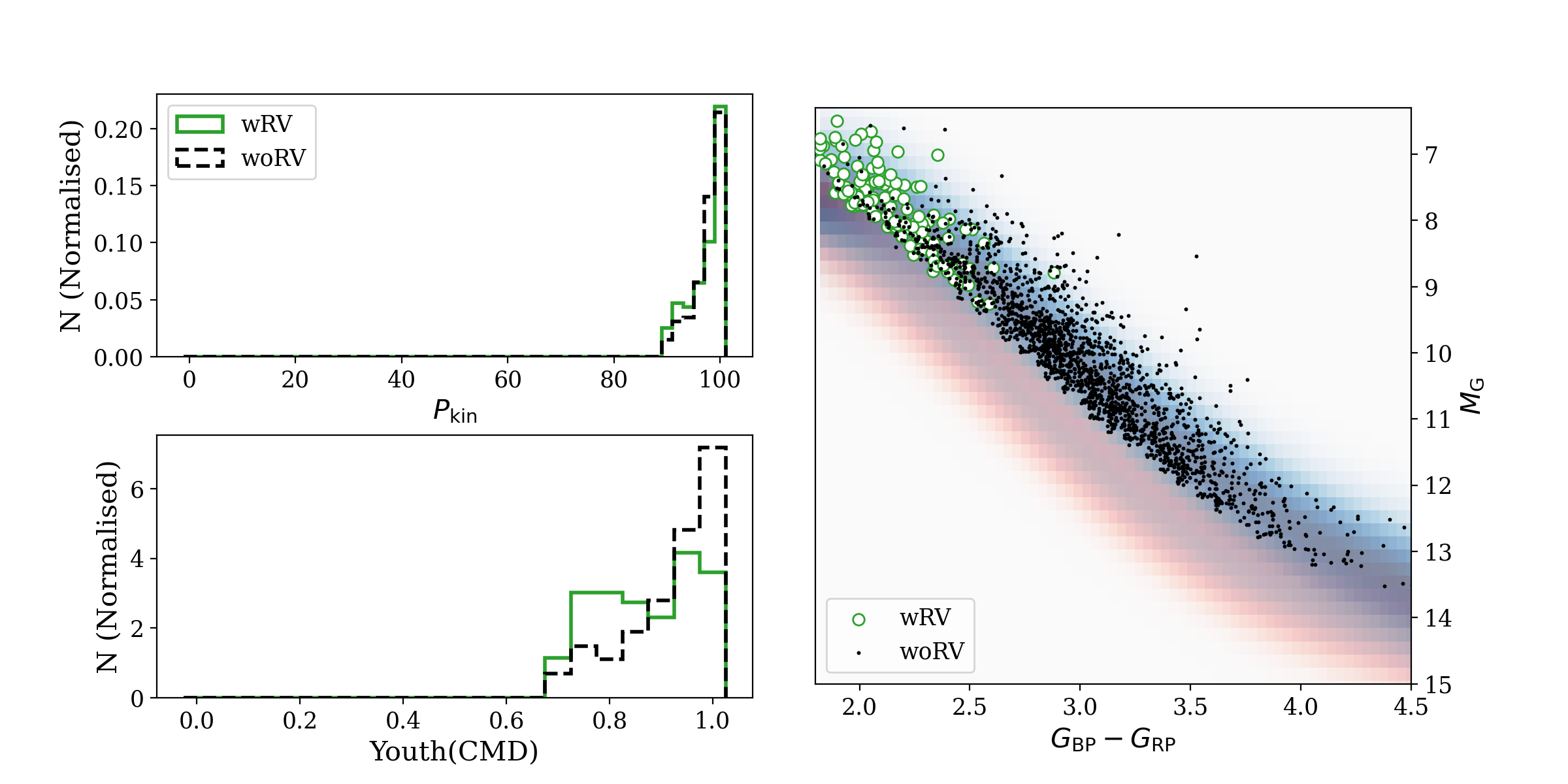}
\caption{Newly identified NYMG candidate members.  Candidates with RV (wRV) and without RV (woRV) are separately displayed.  Left:  histograms of \pkin\ (top) and \ycmd\ (bottom) (normalized; $\Sigma$area = 1.0).
Right: a CMD of these candidates overlaid on the model map identical to Fig.~\ref{fig:ymap4}. }
\label{fig:new_candidates}
\end{figure*}

Histograms of \pkin\ and \ycmd, and a CMD of these 2,002 new members are presented in Fig.~\ref{fig:new_candidates}.  The 139 highly likely members (with RVs) have \bprp\ down to $\sim$2.9 while the other candidates (without RVs) reach down to $\sim$4.5.

Candidates evaluated as young in the $NUV$ youth criterion [\ynuv=1] are almost certainly young NYMG members. Table~\ref{tab:new_nuvlx} shows the number of candidates which satisfy the supplementary youth criterion. About 26 per cent ($N$=522) of the new cool dwarf candidates satisfy the \ynuv\ criterion. For membership confirmation of these candidates, spectroscopic observations are required, especially for those candidates without known RVs. In addition, spectroscopic signs of youth such as strong Li absorption, H$\alpha$ emission, and low surface gravity can confirm their youth.

\begin{table*}
\centering
\begin{threeparttable}
\scriptsize
\setlength\tabcolsep{3pt} 
\begin{tabular}{*{17}{c}}
\hline
{\it Gaia} EDR3 source\_id& Group & \ra\ & \dec\  &  \pmra\ & \pmdec\ & \plx  & $G$ & $J\tnote{a}$ & $H\tnote{a}$ & $K\tnote{a}$ & $W_1\tnote{b}$ & $W_2\tnote{b}$ & $\sum{P_{\rm kin}}$&\pkin & $NUV$ & \ynuv\tnote{c} \\
& & hh:mm:ss & dd:mm:ss & mas yr$^{-1}$ & mas yr$^{-1}$ & mas & mag & mag & mag & mag & mag & mag & \% & \% & mag &  \\ \hline
 2873443317000040832 &      ABDor &  00:02:18.85 &  +30:53:47.0 &   70.71 &  -77.48 &   21.79 & 20.27 & 15.40 & 14.38 & 13.93 & 13.43 & 13.21 &  93 &  92 & 22.39 & 1   \\ 
 4995662902213011712 &      ABDor &  00:04:50.05 &  -42:41:37.6 &  169.76 &   -4.29 &   23.10 & 18.35 & 13.98 & 13.30 & 12.83 & 12.60 & 12.34 &  96 &  96 &  -- & 0   \\ 
 2365527081644959360 &      ABDor &  00:06:34.63 &  -20:27:00.4 &  102.71 &  -25.95 &   17.82 & 18.45 & 14.16 & 13.47 & 13.12 & 12.86 & 12.60 &  94 &  94 &   -- & 2   \\ 
 2876239130257160960 &      ABDor &  00:15:44.85 &  +35:15:58.5 &   54.51 & -257.26 &   58.75 & 18.67 & 13.88 & 12.89 & 12.26 & 11.79 & 11.54 &  98 &  98 & 22.89 & 1   \\ 
 4995583325059035008 &      ABDor &  00:16:59.84 &  -40:56:53.6 &  213.18 &   17.54 &   28.82 & 20.55 & 15.32 & 14.21 & 13.43 & 12.83 & 12.45 &  97 &  97 &  -- & 2   \\ 
\multicolumn{17}{c}{\dots} \\
\multicolumn{17}{c}{\dots} \\
\multicolumn{17}{c}{\dots} \\
\hline
\end{tabular}
\begin{tablenotes}
\item[a] All values are taken from the 2MASS catalogue \citep{cut03}.
\item[b] All values are taken from the ALLWISE catalogue \citep{cut14}.
\item[c] 0: old, 1: young, 2: indecisive of youth; the youth was evaluated using $M_{\rm NUV}$ (see Section 2.2.1 for details).
\end{tablenotes}
\end{threeparttable}
\caption{Newly suggested ultracool dwarf candidate members of NYMGs  identified in this study. The entire list is available online.}
\label{tab:bdcandi}
\end{table*}

\subsubsection{Ultracool dwarf members}

About two thousand ultracool dwarf candidates (\mg$\geq$14; $N$=1,857) have spatio-kinematic membership probabilities larger than 90 per cent.  As explained in Section 2.2.2, the high-sigma astrometric-photometric outliers were eliminated, and 817 objects were retained. Then the criteria for youth evaluation using the three CMDs ($G-J$ vs $M_G$, $J-K$ vs $M_{W1}$, and $H-W_2$ vs $M_{W2}$) were applied to these objects. The numbers of objects assessed as {\it young} and {\it probably young} (See Section 2.2.2) are 180 and 107, respectively. They are marked as blue and cyan dots in Fig.~\ref{fig:bd_candidates}.

Among these 287 candidate members, 56 have been claimed in the literature.  The 231 new ultracool dwarf candidate members are listed in Table~\ref{tab:bdcandi}. 
When the supplementary youth assessment is applied, 33 objects are NUV-bright [\ynuv=1] (See Table~\ref{tab:new_nuvlx}).

Spectroscopic follow-up observations are required to obtain confirmation of youth (e.g., low surface gravity) for the ultracool dwarf candidates and remove any remaining contaminants from the sample.

\subsection{Pilot spectroscopic observations}

\begin{table*}
\scriptsize
\begin{threeparttable}
\setlength\tabcolsep{2pt} 
\begin{tabular}{*{19}{c}}
\hline
{\it Gaia} EDR source\_id & Group & SpT & \ra\ & \dec\  &  \pmra\ & \pmdec\ & \plx & RV &  $G$ & $G_{\rm BP}$ & $G_{\rm RP}$ & $NUV$ & EW(H$\alpha$)\tnote{a} & EW(Li)\tnote{b} & $P_{\rm kin}$ & {\it Y(CMD)}\tnote{c}& {\it Y(NUV)} \tnote{d} & Obs. date \\
& & & hh:mm:ss & dd:mm:ss & mas yr$^{-1}$ & mas yr$^{-1}$ & mas & \kms & mag & mag & mag & mag & \AA & m\AA & \% & & & yyyy-mm-dd  \\ \hline

  4904674604163917696 &      ABDor &         M5 &  00:15:55.57 &  -61:37:51.9 &   68.17 &  -41.96 &   18.21 &   22.60 $\pm$ 0.46 & 13.43 & 14.94 & 12.23 & 21.88 &  6.05$\pm$1.08 &   $<$91 &    98 &  0.81  & 1& 2018-10-28\\  
  2788357364871430400 &      ABDor &         M4 &  01:02:50.98 &  +18:56:54.4 &   97.34 &  -59.77 &   26.19 &   -7.35  $\pm$ 2.17 & 12.58 & 14.27 & 11.33 & 20.14 &  7.65$\pm$1.24 &     28$\pm$14 &    60\tnote{e} &  0.99  &1&  2016-11-16\\  
  3274291270411229056 &      ABDor &         M4 &  03:36:40.84 &  +03:29:19.5 &  122.13 & -120.24 &   37.35 &   28.20 $\pm$  7.90 & 12.39 & 14.15 & 11.12 & 19.90 &  5.71$\pm$0.83 &   $<$76 &   100 &  0.96 & 1&  2016-11-16\\  
  3341248779763320448 &      ABDor &         M2 &  05:40:16.09 &  +12:39:00.8 &  -16.57 & -260.86 &   29.07 &    7.10 $\pm$1.65  & 10.45 & 11.53 &  9.42 &   -- &  2.66$\pm$0.39 &   $<$64 &   100 &  0.90  & 2& 2016-11-15\\  
  5343936233991962880 &      ABDor &         M4 &  11:45:47.73 &  -55:20:30.7 &  -97.61 &    5.96 &   24.19 &   13.57$\pm$ 2.89 & 12.17 & 13.23 & 10.54 & 18.12 &  7.60$\pm$1.15 &  $<$129 &    69\tnote{e} &  0.99& 1&   2016-07-17\\  
   3576519980067360768 &      ABDor &         M5 &  12:26:44.13 &  -12:29:17.5 & -170.94 &  -89.12 &   33.05 &    7.77$\pm$0.62 & 12.10 & 13.47 & 10.81 & 19.50 &  3.65$\pm$0.81 &   $<$54 &   100 &  0.71  & 1& 2017-04-14\\  
  6005261829198023296&      ABDor &         M6 &  15:13:55.55 &  -39:41:13.7 &  -79.18 &  -90.87 &   25.37 &    0.88$\pm$2.58  & 14.04 & 16.29 & 12.66 &   -- & 19.89$\pm$1.91 &    981$\pm$324  &    98 &  1.00   & 0& 2018-06-04\\  
  6724937269577837696 &      ABDor &         M6 &  18:15:30.61 &  -41:06:10.2 &   28.46 &  -95.86 &   25.38 &  -21.87$\pm$0.62 & 13.29 & 14.81 & 11.73 &   -- &  4.49$\pm$1.02 &   $<$68 &    96 &  0.98   & 2& 2018-06-04\\  
 6362966355075756928 &      ABDor &         M4 &  19:52:27.05 &  -77:35:28.4 &   66.19 & -107.26 &   25.69 &   37.05 $\pm$5.82 & 12.49 & 14.00 & 11.29 & 20.26 &  4.06$\pm$0.75 &   $<$79 &    96 &  0.91 &1  & 2016-07-17 \\  
6878315602572236160 &      ABDor &         M4 &  20:03:01.59 &  -14:22:28.3 &   28.24 &  -97.47 &   17.45 &  -10.74$\pm$0.47 & 13.56 & 14.67 & 11.97 & 20.77 &  4.76$\pm$1.00 &   $<$61 &    98 &  0.81  & 1& 2018-10-27 \\  
  6465741903107929472&      ABDor &       M7/8 &  21:38:35.31 &  -50:51:10.8 &   99.56 &  -60.37 &   22.09 &  -13.82 $\pm$ 1.89& 13.92 & 15.67 & 12.59 & 21.76 &  5.97$\pm$1.08 &   $<$89 &    99 &  0.86  & 1& 2018-10-29 \\  
    2701170632570283904 &      ABDor &       M7/8 &  21:39:01.79 &  +07:00:34.4 &   70.24 &  -45.99 &   18.64 &  -14.67$\pm$0.63 & 14.36 & 16.11 & 13.10 & 21.45 &  7.82$\pm$1.43 &   $<$93 &    97 &  0.78  &1&  2018-10-27\\  
  2680136803331701504 &      ABDor &         M3 &  21:55:17.39 &  -00:46:23.2 &   63.58 &  -55.53 &   17.89 &  -11.84$\pm$0.43 & 14.10 & 15.87 & 12.83 & 21.19 & 12.36$\pm$1.71 &   $<$76 &    97 &  0.96  & 1& 2017-09-11 \\  
  6820341000133968768 &      ABDor &         M5 &  22:11:42.09 &  -20:44:18.1 &  146.52 &  -64.52 &   24.15 &   -9.41$\pm$ 1.44& 12.49 & 13.98 & 11.30 & 19.88 &  4.09$\pm$0.81 &   $<$95 &   100 &  0.85  & 1& 2018-10-29 \\  
   5794808569134876416 &      Argus &         M4 &  15:36:29.97 &  -72:20:20.1 &  -55.30 &  -71.20 &   16.56 &   -5.25 $\pm$ 0.31 & 12.63 & 13.96 & 11.48 &   -- &  3.95$\pm$0.91 &   $<$61 &    81\tnote{e} &  0.93& 2  & 2017-04-14 \\  
 6316591222362496  &       BPMG &         M5 &  02:48:16.81 &  +05:37:06.0 &   74.49 &  -53.62 &   21.45 &    8.24 $\pm$1.09  & 13.14 & 14.91 & 11.87 & 20.20 &  8.31$\pm$1.32 &   $<$34 &    96 &  1.00 & 1&  2018-10-30\\  
  9124679495764864 &       BPMG &         M6 &  03:23:39.16 &  +05:41:15.3 &   83.32 &  -69.50 &   26.64 &   10.12$\pm$0.62  & 13.00 & 14.79 & 11.73 & 21.02 &  6.76$\pm$1.18 &    156$\pm$92&    90 &  0.99  & 1& 2018-10-30 \\  
   3290083835095329408  &       BPMG &         M4 &  05:07:49.14 &  +08:29:37.0 &    4.75 &  -63.88 &   17.91 &   10.65 $\pm$1.58& 13.04 & 14.46 & 11.80 &   -- &  2.45$\pm$0.63 &   $<$61 &    92 &  0.94   & 0&2018-10-30 \\  
  6455829427826823168 &       BPMG &         M7 &  20:44:48.19 &  -58:03:28.8 &   39.32 &  -56.76 &   16.07 &    1.51  $\pm$1.16& 14.47 & 16.47 & 13.14 & 22.00 & 11.80$\pm$1.54 &    846$\pm$251 &    99 &  1.00   &1& 2018-10-29 \\  
 2594209965325689600 &       BPMG &         M6 &  22:29:44.26 &  -18:23:13.4 &  110.25 &  -59.18 &   27.01 &   -3.86 $\pm$2.04& 13.84 & 15.66 & 12.57 & 21.72 &  5.30$\pm$1.13 &  $<$102 &   100 &  0.80  & 1& 2018-10-26 \\  
  5093056748849753344 &     Carina &         M4 &  04:20:36.75 &  -18:53:08.3 &   30.88 &  -11.80 &   10.18 &   20.16$\pm$ 1.00 & 13.82 & 15.33 & 12.61 & 21.25 &  5.54$\pm$1.09 &   $<$59 &   100 &  0.99  & 1& 2018-10-29 \\    
  2977580846408038016 &     Carina &       M7/8 &  04:46:52.77 &  -19:36:45.4 &   29.28 &   -9.13 &   10.69 &   19.11 $\pm$0.56 & 14.97 & 16.87 & 13.67 & 21.76 &  9.67 $\pm$1.22 &   $<$71 &    99 &  1.00  & 1& 2018-10-30 \\  
4880597979576260480 &     Carina &       M7/8 &  04:53:54.38 &  -27:05:46.5 &   28.91 &   -2.16 &   10.77 &   27.39 $\pm$1.54 & 15.89 & 17.82 & 14.58 & 23.09 &  9.29$\pm$1.34 &  $<$355 &    94 &  0.92   &1& 2018-10-29 \\  
3199199681413101056 &    ThOrCol &         M6 &  04:22:12.00 &  -06:30:19.5 &   39.23 &  -28.55 &   13.61 &   22.89$\pm$ 0.93& 14.47 & 16.20 & 13.21 &  21.12 &  6.86$\pm$1.11 &   $<$74 &    71\tnote{e} &  0.97  &1&  2018-10-29 \\  
3226167471826076672 &    ThOrCol &         M4 &  04:48:22.55 &  -01:53:56.0 &   27.14 &  -64.92 &   14.18 &   23.96$\pm$ 0.57 & 13.99 & 15.59 & 12.75 &   -- &  8.57$\pm$1.30 &   $<$72 &   100 &  0.98  & 0& 2018-02-03 \\  
   2883627990131188352 &    ThOrCol &         M4 &  05:57:29.90 &  -38:43:03.7 &   14.69 &   10.13 &   10.66 &   18.07 $\pm$0.51 & 13.83 & 15.30 & 12.63 & 19.94 &  8.67$\pm$1.36 &   $<$66 &    92 &  0.99  &1&  2017-09-14\\  
  2916409310839890176 &    ThOrCol &         M4 &  05:59:41.10 &  -23:19:09.3 &    6.40 &   -5.61 &   13.93 &   27.25 $\pm$ 4.10 & 13.42 & 15.01 & 12.19 & 20.38 & 12.07$\pm$1.32 &   $<$83 &   100 &  1.00 & 1&  2016-11-15\\  
 5213914375483588480&    ThOrCol &         M4 &  08:10:36.42 &  -74:58:12.3 &  -37.34 &   56.34 &   12.79 &   15.42 $\pm$0.42 & 13.08 & 14.45 & 11.92 & 20.52 &  5.08 $\pm$0.91&   $<$59 &    90 &  0.98  & 1& 2018-02-03 \\  
  5806701677174982912 &    ThOrCol &         M4 &  16:14:40.97 &  -72:02:24.3 &  -38.92 &  -69.57 &   14.72 &    9.37$\pm$ 0.85 & 13.70 & 15.19 & 12.47 &   -- &  5.29$\pm$1.02 &   $<$52 &    84\tnote{e} &  0.91  &0&  2018-05-05 \\  
2466437475703854208  &  TucCol &         M4 &  02:01:50.73 &  -07:26:58.2 &   95.62 &  -51.87 &   22.18 &    6.46 $\pm$ 2.51& 14.07 & 15.86 & 12.80 & 22.06 &  9.43$\pm$1.44 &  $<$151 &    99 &  0.86  & 1& 2016-11-16 \\  
4946915641880235904&  TucCol &         M5 &  02:39:48.29 &  -42:53:05.0 &  104.39 &  -25.33 &   25.31 &    9.95$\pm$ 1.89 & 12.38 & 14.09 & 11.11 &  -- & 10.36$\pm$1.38 &   $<$33 &   100 &  1.00  & 0& 2018-10-26 \\  
  3176591214084032128 &  TucCol &       M1/2 &  04:12:55.75 &  -14:18:59.0 &   60.75 &  -25.35 &   17.38 &   13.78 $\pm$ 1.53 & 11.90 & 12.99 & 10.86 & 19.80 &  2.21$\pm$0.54 &     52$\pm$24 &   100 &  0.76  & 1& 2017-09-14 \\  
6375044288444444672 &  TucCol &         M5 &  21:07:22.49 &  -70:56:12.6 &   26.72 &  -92.28 &   20.70 &    5.70 $\pm$0.61& 12.60 & 14.06 & 11.41 & 19.47 &  6.18$\pm$1.16 &   $<$54 &   100 &  0.97 & 1&  2018-10-27 \\  
 2312002988748576640 &  TucCol &         M2 &  23:43:26.82 &  -34:46:57.8 &   91.78 &  -76.92 &   26.03 &   -2.34 $\pm$ 0.81& 11.01 & 12.12 &  9.92 & 19.17 &  2.49$\pm$0.55 &   $<$40 &   100 &  0.83  & 1& 2018-10-26\\  
  6534521406305916928 &  TucCol &        M6 &  23:45:25.66 &  -40:20:15.0 &  108.41 &  -78.84 &   26.30 &    6.57 $\pm$2.47 & 13.64 & 15.41 & 12.37 & 21.39 &  6.46$\pm$1.13 &   $<$62 &    56\tnote{e} &  0.78  & 1& 2018-10-26\\  
  \hline
\end{tabular}
\begin{tablenotes}
\item[a] Emissions. Uncertainties are calculated using Eqn. 6 of \citet{cay88}.
\item[b] In the case of non-detection, upper limits are provided. Uncertainties are calculated using Eqn. 6 of \citet{cay88}.
\item[c] \ycmd.
\item[d] \ynuv.
\item[e] While the candidate's \pkin\ of a single group is less than 90 per cent, the combined \pkin\ of groups ($\Sigma$\pkin) is larger than 90 per cent.  The candidate has an ambiguous status being a member of a single group.
\end{tablenotes}
\end{threeparttable}
\caption{Newly confirmed NYMG members after obtaining spectra.}
\label{tab:newbonafide2}
\end{table*}

\begin{figure}
\begin{subfigure}[b]{.99\linewidth}
\includegraphics[width=0.95\linewidth]{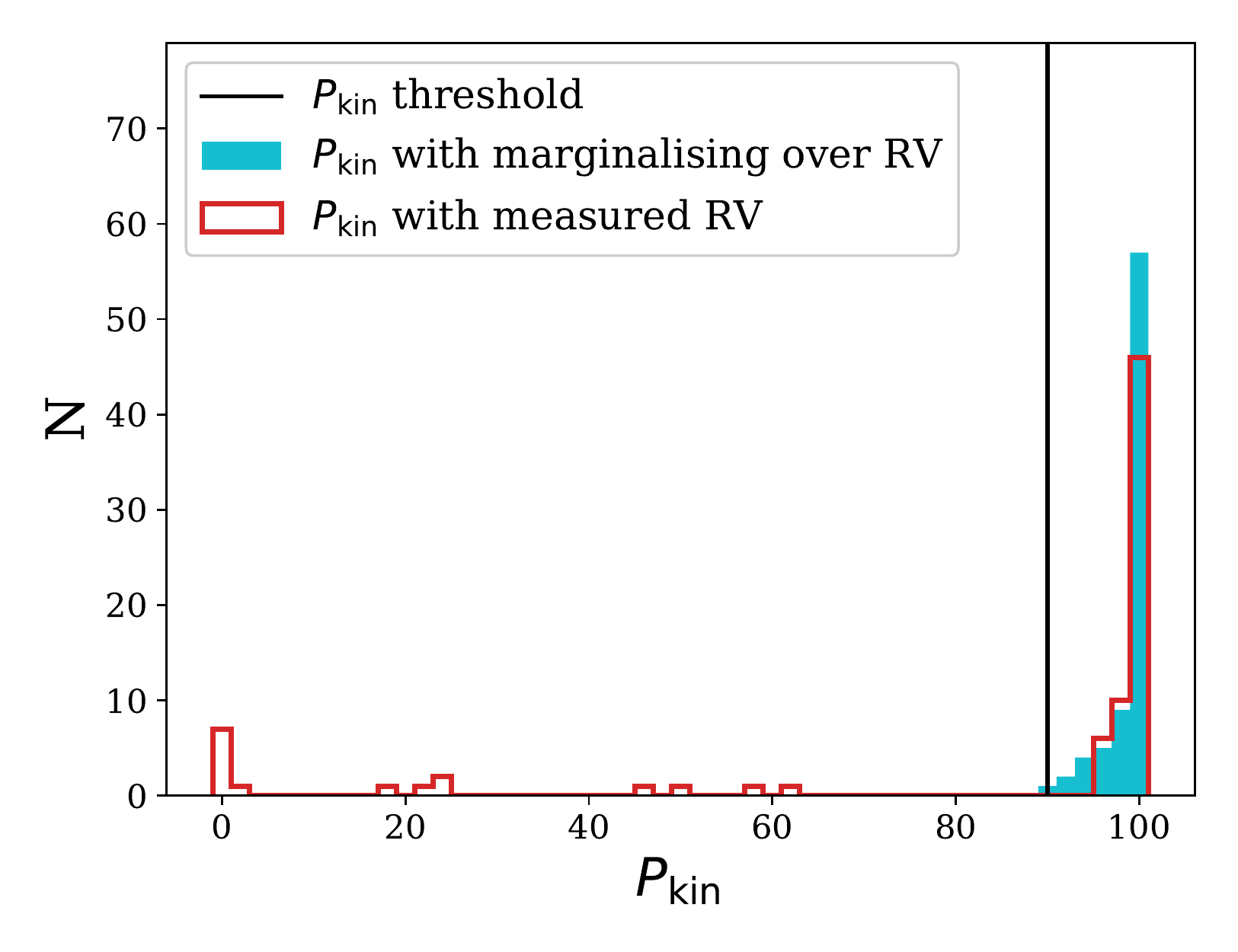}
\end{subfigure}
\begin{subfigure}[b]{.99\linewidth}
\includegraphics[width=0.95\linewidth]{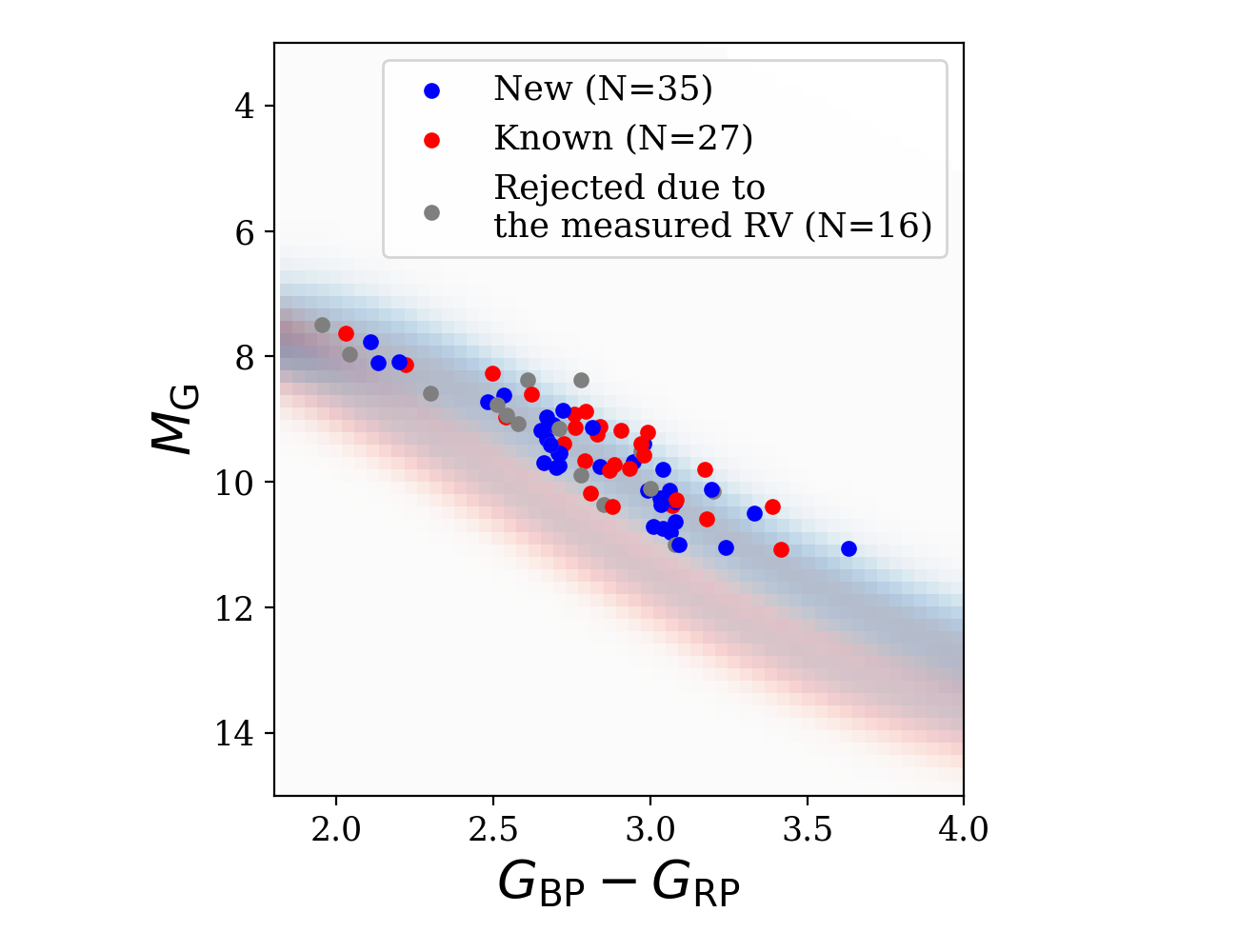}
\end{subfigure}
\caption{The spectroscopically observed NYMG candidate members.
Top: histograms of \pkin\ values calculated marginalising over RV (cyan bars) and with the measured RV (empty bars). The vertical line indicates the selection threshold of spatio-kinematic candidates (\pkin=90 per cent). 
Bottom: the \bprp\ vs \mg\ CMD. All the candidates are bright on the CMD (\ycmd$\geq$0.7). Newly confirmed members, previously claimed members from the literature, and the rejected candidates are marked as blue, red, and grey circles, respectively. }
\label{fig:sso1}
\end{figure}

\begin{figure*}
\includegraphics[width=0.95\linewidth]{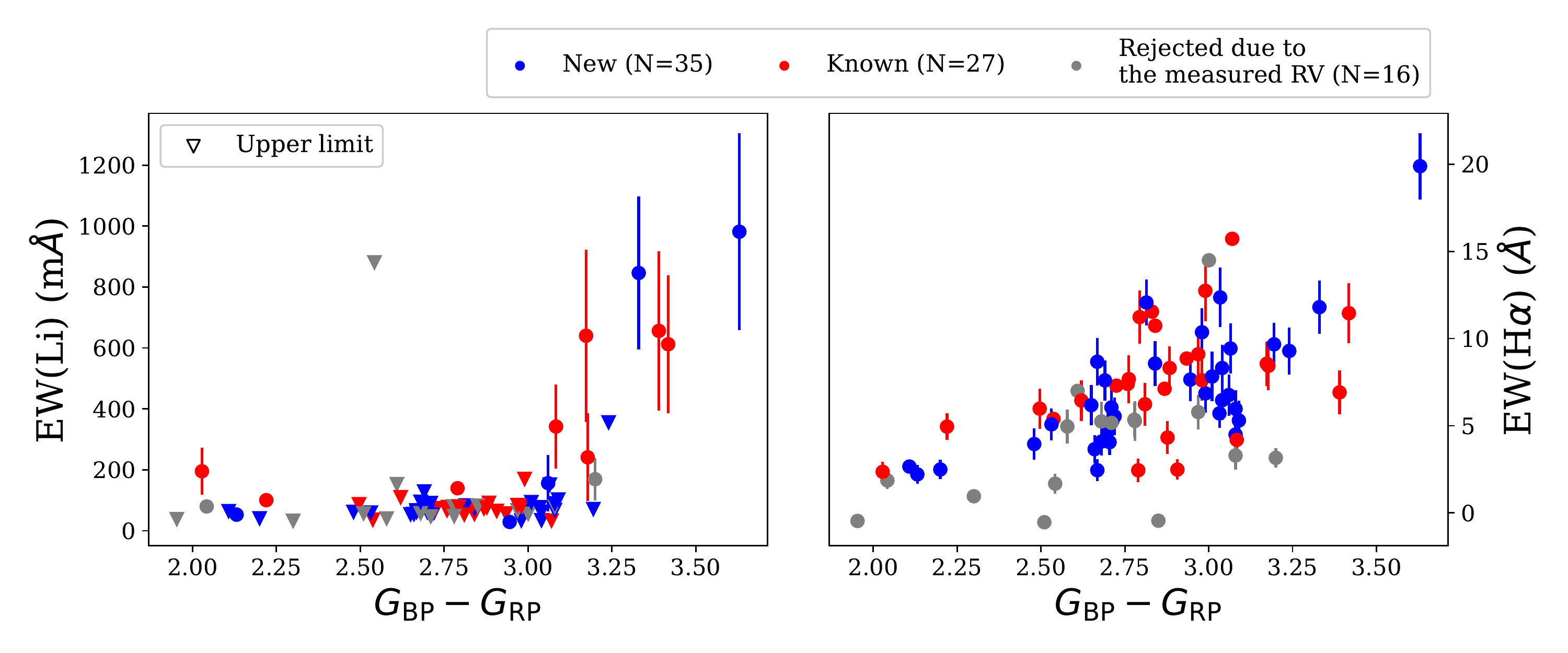}
\caption{The spectroscopically observed NYMG candidate members. Newly confirmed members, previously claimed members from the literature, and the rejected candidates are marked as blue, red, and grey circles, respectively. Left: equivalent widths of the Li absorption line at 6708\AA. Right: equivalent widths of the H$\alpha$ emission line.}
\label{fig:sso2}
\end{figure*}

 We obtained optical spectra ($\lambda/\Delta\lambda\sim$7000) for a subset of the candidates with the Wide Field Spectrograph (WiFeS; \citealt{dop07}) on the ANU 2.3-m telescope at Siding Spring Observatory.
We used the red R7000 grating, which covers 5300 to 7060\AA.
We measured the RV of each spectrum with \textsc{ispec} via cross-correlation against the provided HARPS/SOPHIE line mask  \citep{bla14, bla19}.
Equivalent widths of Li 6708\AA\ and H$\alpha$ were measured with the splot task in \textsc{iraf}. \newline

Among $\sim$2000 candidate NYMG members, we have observed $>$100 sources with WiFeS since 2016.  Because our BAMG scheme has been actively developed during this time as well, some observed candidate members would not have been retained purely based on the selection criteria described in earlier sections.
Seventy eight candidates fulfilling the \pkin\ and \ycmd\ selection criteria were observed with WiFeS.  Their newly obtained RVs were used in re-calculation of their \pkin, and 79 per cent ($N$=62) still return \pkin\ above the threshold. Such a large hit rate proves the effectiveness of the BAMG code in the NYMG member identification.
Among these 62 members, 27 were claimed previously in the literature.  Excluding these 27 members, the newly confirmed NYMG members ($N$=35) are listed in Table~\ref{tab:newbonafide2}.

Fig.~\ref{fig:sso1} displays histograms of \pkin\ and a CMD for the 78 sources, including the 16 objects rejected as their velocities are inconsistent with membership in the proposed NYMG.  Equivalent widths of their Li and H$\alpha$ lines are presented in Fig.~\ref{fig:sso2}.
Even though these values are inconclusive to evaluate an object's youth in the low-mass regime, most have strong H$\alpha$ emission, corroborating their CMD youth.

\clearpage

\section{Discussion}

\subsection{Population statistics and survey limit}

\begin{figure*}
\includegraphics[width=0.99\linewidth]{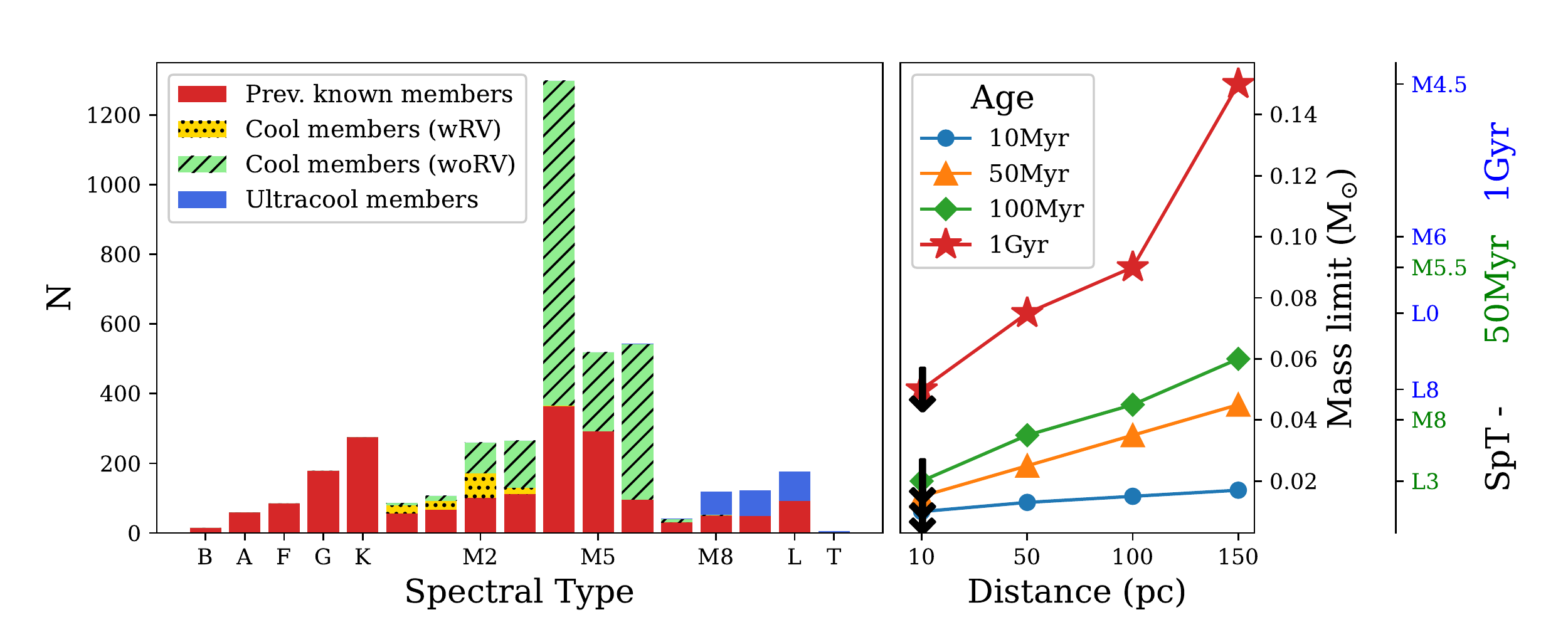}
\caption{Left: spectral types of previously known members from the literature (See Section 3) and new candidate members from this study. Since there is a large number of M types, the M types are split into subtypes for better display.
Right: the mass limit of this study assuming that the limiting magnitude of {\it Gaia} is $G$=20. For each age, the \mg\ is converted into mass and effective temperature using theoretical isochrones by \citet{bar15}. The effective temperatures are converted into spectral types using a conversion relation by 
Mamajek\textsuperscript{3} and \citet{pec13}.
Spectral types for 50 Myr and 1 Gyr are presented on the right side of the y-axis. The downwards arrows show the upper limit as the theoretical isochrones have the lower mass limit in their calculation.}
\label{fig:spt}
\end{figure*}

In this study, we have identified 2,002 cool dwarf and 231 ultracool dwarf candidate NYMG members. 
Their spectral types were estimated using theoretical isochrones (BHAC15; \citealt{bar15}) as follows.
The NYMG membership provides an age for each source based on that of the designated group (TWA: 10, BPMG: 20, ThOrCol, TucCol, \& Carina: 30, Argus: 40, ABDor \& VCA:100 Myr). The isochrone model converts a broadband colour to an effective temperature. The colours used in the conversion are \bprp\ and $G-J$ for cool and ultracool dwarfs, respectively. An effective temperature is converted to a spectral type using a sequence by Mamajek\footnote{http://www.pas.rochester.edu/$\sim$emamajek  /EEM\_dwarf\_UBVIJHK\_colors\_Teff.txt} and \citet{pec13}.
Fig.~\ref{fig:spt} (left) illustrates the distribution of spectral types of new candidate members from this study and those of previously known members. For previously known members, spectral types from the literature were adopted. New highly likely members with RVs are mainly  early-M types while the majority of those missing RVs are mid-M types. The number of new ultracool dwarf candidates is similar to that of the previously known ultracool dwarf members, which would double the ultracool dwarf NYMG sample if all of their memberships are confirmed.

Our survey for the lowest-mass NYMG members is incomplete due to {\it Gaia} EDR3  only being sensitive out to 20-30 pc for mid-L members \citep{sch20}.  The completeness of individual NYMGs must differ as their mean distances and ages vary widely.
Fig.~\ref{fig:spt} (right) illustrates the lowest mass detected by {\it Gaia} assuming a limiting magnitude of $G$=20. The calculation is based on theoretical isochrones by \citet{bar15}. Obviously, the younger, hence hotter objects have lower mass detection limits than their older counterparts. In the nearest $\sim$10 pc, {\it Gaia} EDR3 is complete down to late-L regardless of age. At 50 pc, $\sim$L3 and $\sim$L0 are the latest detectable spectral types for $<$50 Myr and $\sim$Gyr objects, respectively. When considering mass instead of spectral type at 50 pc, detection limits for young ($<$50 Myr) and old ($\sim$Gyr) objects are 0.025 \msun\ and 0.075 \msun, respectively.
At 150 pc, late-M types (0.02$-$0.06 \msun) appear to be the faintest available NYMG members, while mid-M stars (0.15 \msun) are the lowest mass old field objects.
Therefore, in this study, most M type NYMG members were likely to be found, and some nearby ($\lesssim$50 pc) L type members were also identified. As mentioned in Section 3.1, the recovery rate of our analysis is $\sim$56 per cent when compared to the literature. The true completeness must be (much) lower than that value because of (1) the incompleteness of the NYMG census in the literature, (2) the survey limit of {\it Gaia}, and (3) hidden/uncovered NYMG groups within 150 pc. In spite of the incompleteness, the newly identified low-mass members from this study  significantly contribute to the full census of NYMG members.

\subsection{Age of the $\beta$ Pictoris Moving Group (BPMG)}

\begin{figure*}
\includegraphics[width=0.95\linewidth]{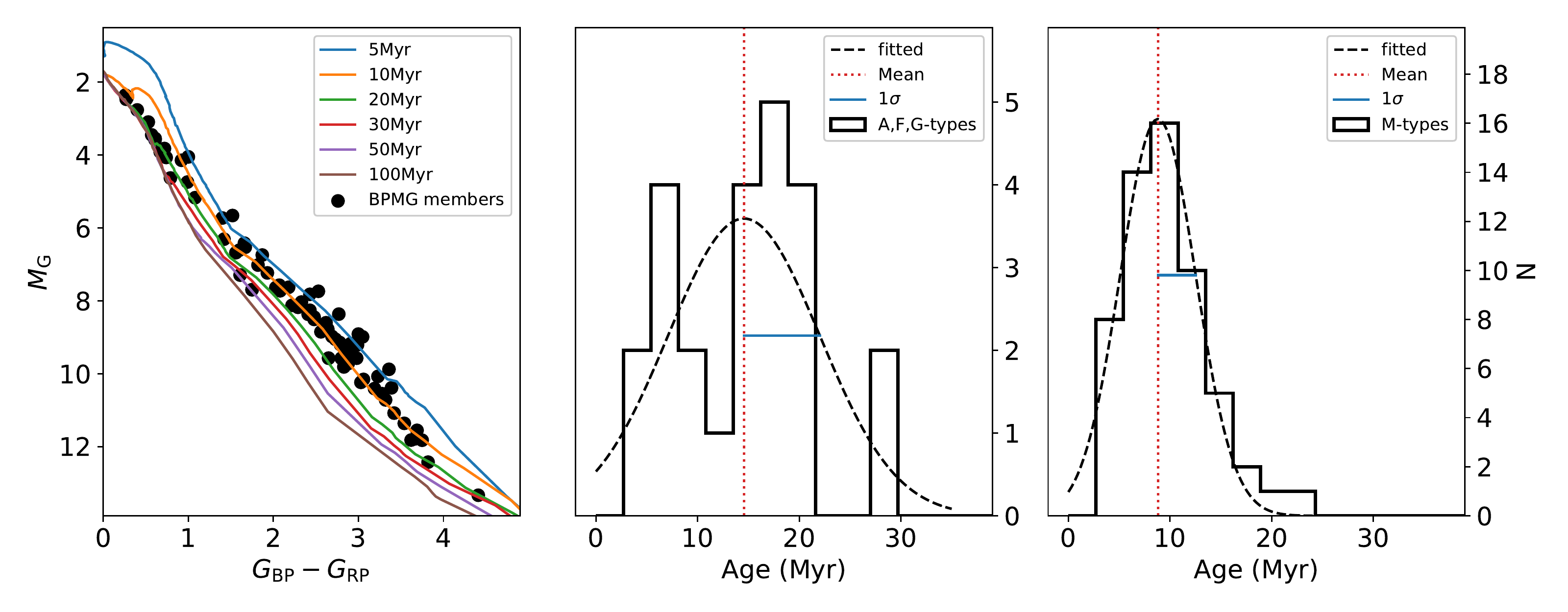}
\caption{Age estimation for the BPMG.  Left: a colour-magnitude diagram for BPMG bona fide members from Paper II and newly confirmed members from this study.  Combined isochrones (PARSEC-COLIBRI at $\geq$0.7\msun\ and BHAC15 at $<$0.7\msun, see text) are used to interpolate the age of each member.
Middle and Right: histograms of the estimated ages of BPMG members using A, F, and G types (middle) and M types (right). The mean ages using the Gaussian fitting are 14.6 Myr and 8.8 Myr for AFG-types and M-types, respectively. When all members are used together, the mean age is 9.4 Myr.}
\label{fig:bpmgage}
\end{figure*}

The estimated age of the BPMG has a wide range (12$-$25 Myr; \citealt{zuc01, son03, bin14, mam14, bel15}) caused by different selection of input data, and/or method dependency.  Later analyses \citep{bin14, mam14, bel15} generally find older ages ($>$20 Myr).
\citet{mam14} find ages (20$-$22 Myr) using multiple isochrone models [Yonsei-Yale; \citet{dem04}, Dartmouth; \citet{dot08}, Pisa; \citet{tog11}, and PARSEC; \citet{bre12}].
However, the members used in their analyses have spectral types in the range of A0 to G9, whose over-luminosity compared to old MS stars is smaller than those of M-type stars.
Therefore, including lower-mass members can yield a more robust estimate of the group age.
In this re-analysis, we adopted two isochrones: PARSEC-COLIBRI \citep{mar17} and BHAC15 \citep{bar15}.

Following the scheme of \citet{gol18}, we used BHAC15 for intermediate-mass stars down to the ultracool dwarfs and PARSEC-COLIBRI for the higher-mass stars.
We adopted a boundary of 0.7\msun\ between the two models \citep{gol18}.
The age of each source is estimated via linear interpolation using the $G\rm_{BP}$, $G\rm_{RP}$, and $G$ magnitudes. In the case of known and unresolved binaries, the source is split into two by adding 0.7 mag in $G\rm_{BP}$, $G\rm_{RP}$, and $G$ assuming they are equal luminosity binaries.  

Fig.~\ref{fig:bpmgage} presents a CMD and distribution of the estimated ages of BPMG members.  The distribution of ages is fitted with a Gaussian function to evaluate the age of the BPMG.
When the entire BPMG membership is used, the fitting has a mean value of 9.4 Myr with a standard deviation ($\sigma$) of 4.9 Myr. This age is significantly younger than the currently accepted age of BPMG ($\sim$20 Myr).  If only lower-mass members (\bprp$\geq$1.8) are used, the estimated age becomes younger $\sim$8.8 Myr ($\sigma$=3.7 Myr).  When the high mass members (\bprp$<$ 1.8) are used, the estimated age is still young ($\sim$14.6 Myr, $\sigma$=7.5 Myr).  This systematic bias of higher mass stars appearing older than low mass stars on theoretical isochrones is a well-known effect \citep{nay09, her15}. Even though there is a mass-dependency issue in the age evaluation, the BPMG members overall seem to be younger than the commonly adopted age of $\sim$20 Myr.

This situation is similar to the case of the Lower Centaurus Crux (LCC) subgroup of the nearby Scorpius-Centaurus OB association.
\citet{mam02} and \citet{pec16} determined isochronal ages of LCC as 16 Myr based upon main-sequence turn-off, G-type and F-type members. The ages based upon the massive members are significantly older than those from \citet{gol18} and \citet{son12}.  \citet{gol18} estimated the isochronal ages of subgroups of LCC to be 7-10 Myr using high- to low-mass members (B-type stars to ultracool dwarfs).  \citet{son12} assessed the age of LCC as 10 Myr from a Li $\lambda$6708 absorption strength analysis (also see Murphy, Lawson \& Bento, 2015). 
Although this well-known discrepancy between the isochronal ages of massive and low-mass stars is yet to be resolved, there are some possible explanations regarding low-mass stars.
For instance, standard stellar evolutionary models (e.g., \citealt{bar98, tog11}) do not include accretion, rotation, or magnetic fields. Rotation and magnetic fields  slow the contraction of T-Tauri stars, inducing a larger radius and/or a cooler surface temperature at a given age \citep{fei16, ama19}.  Episodic accretion, although a very rare event and effective only within 10 Myr age \citep{bar10}, also can cause young stars to be very luminous. 
In contrast, scattering, disks, or star spots reduce stellar luminosities \citep{jac09, dun10}. This apparent over-luminosity of young M-type stars compared to theoretical expectations can therefore, be attributed to the combination of these effects. 
Keeping this in mind, the CMD age of the BPMG, mostly influenced by low mass members, is $\sim$10 Myr.

\section{Conclusion and Summary}

In this study, we have identified new low-mass (M0 to mid-L) NYMG members from {\it Gaia} EDR3.  Two procedures were used in the identification: (1) spatio-kinematic membership probability calculation via BAMG and (2) photometric assessment of youth.

For cool dwarfs (\mg$\leq$14 mag), the youth assessment was performed primarily using the location on the \bprp\ vs \mg\ diagram.  After removing known members, $\sim$2000 new cool dwarf candidate members were identified, including 139 stars with RVs.  This subset of candidate members have full astrometric data which gives a more rigorous spatio-kinematic membership assessment.  Although there is a possibility that these stars are multiple systems which elevate them on the CMD and cause them to be falsely assessed as young, we call them highly likely new NYMG members.

For identifying ultracool dwarf members (\mg$>$14 mag), youth was assessed using three CMDs ($G-J$ vs \mg, $J-K$ vs $M_{W1}$, and $H-W_2$ vs $M_{W1}$).  After cross-matching with previously known ultracool dwarf NYMG members, 231 new candidate members were identified. \newline
Although our survey is incomplete due to the sensitivity of {\it Gaia}, the spectral type distribution of NYMG members shows that our survey significantly contributes to the identification of low-mass NYMG members. 
The supplementary youth criterion using $NUV$ selects more certain young candidates ($\sim$500 and $\sim$40 for cool and ultracool dwarfs, respectively).

Finally, a new age estimation of the $\beta$ Pictoris Moving Group (BPMG) using the updated list of bona fide members and {\it Gaia} EDR3, results in a younger age ($\sim$10 Myr) compared to the previously derived ages ($\sim$20 Myr). Although outside the scope of this work, it would be worthwhile to re-calculate the Lithium Depletion Boundary age of the BPMG with the updated membership list.

As the main results of this study, we provide lists of spectroscopically confirmed members,  highly likely members, and candidate members of NYMGs.

\section*{Acknowledgements}

We acknowledge the important and thorough and valuable review by the referee that improved the manuscript significantly.

This research was supported by Basic Science Research Program through the National Research Foundation of Korea (NRF) funded by the Ministry of Education (No. NRF-2021R1I1A1A01047972).

This work has made use of data from the European Space Agency (ESA) mission {\it Gaia} (\url{https://www.cosmos.esa.int/gaia}), processed by the {\it Gaia} Data Processing and Analysis Consortium (DPAC, \url{https://www.cosmos.esa.int/web/gaia/dpac/consortium}). Funding for the DPAC has been provided by national institutions, in particular the institutions participating in the {\it Gaia} Multilateral Agreement.

\section*{Data availability}

The data underlying this article are available in the article and in its online supplementary material.








\appendix

\section{Generation of \ycmd\ map}

\begin{figure*}
\includegraphics[width=0.9\linewidth]{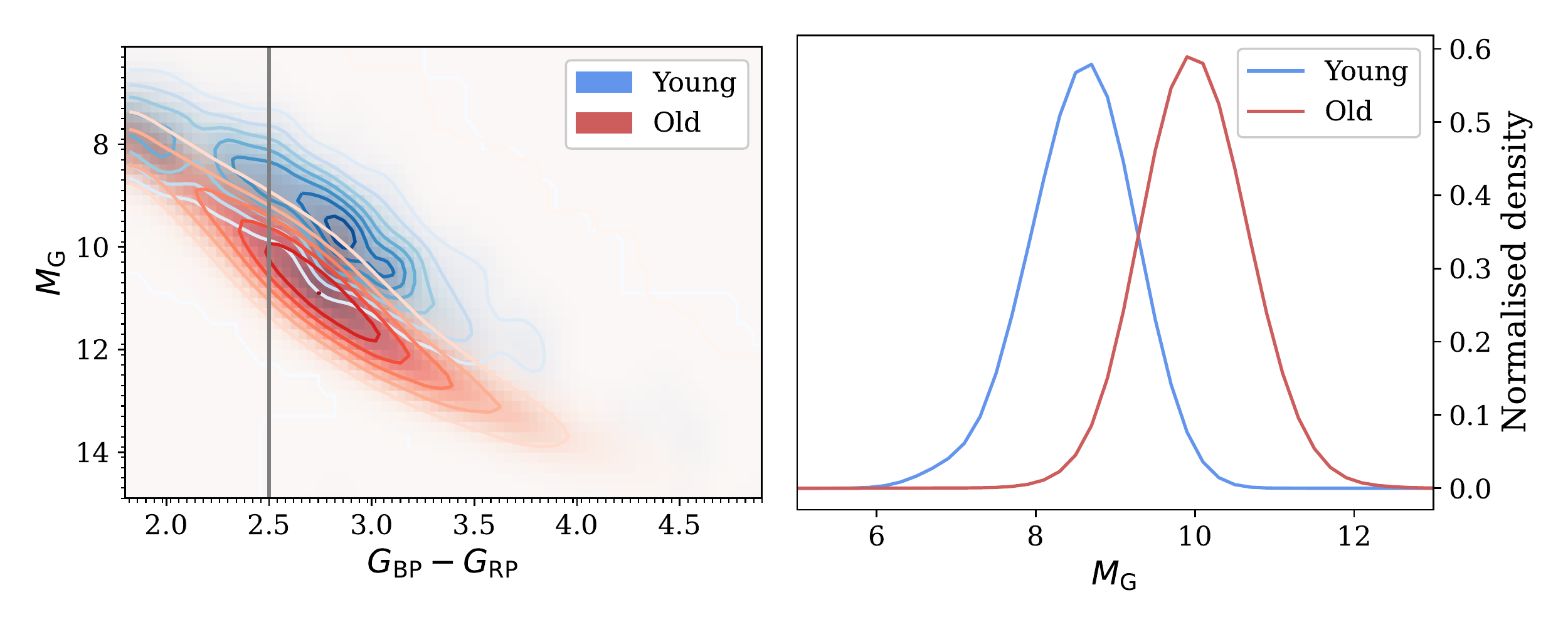}
\caption{Left: smoothed density maps after applying a gaussian filter ($\sigma$=2.0).
Blue and red maps are for young and old stars, respectively.
Right: density distributions obtained by slicing the left panel at \bprp\ = 2.5.}
\label{fig:ymap2}
\end{figure*}

\begin{figure*}
\includegraphics[width=0.45\linewidth]{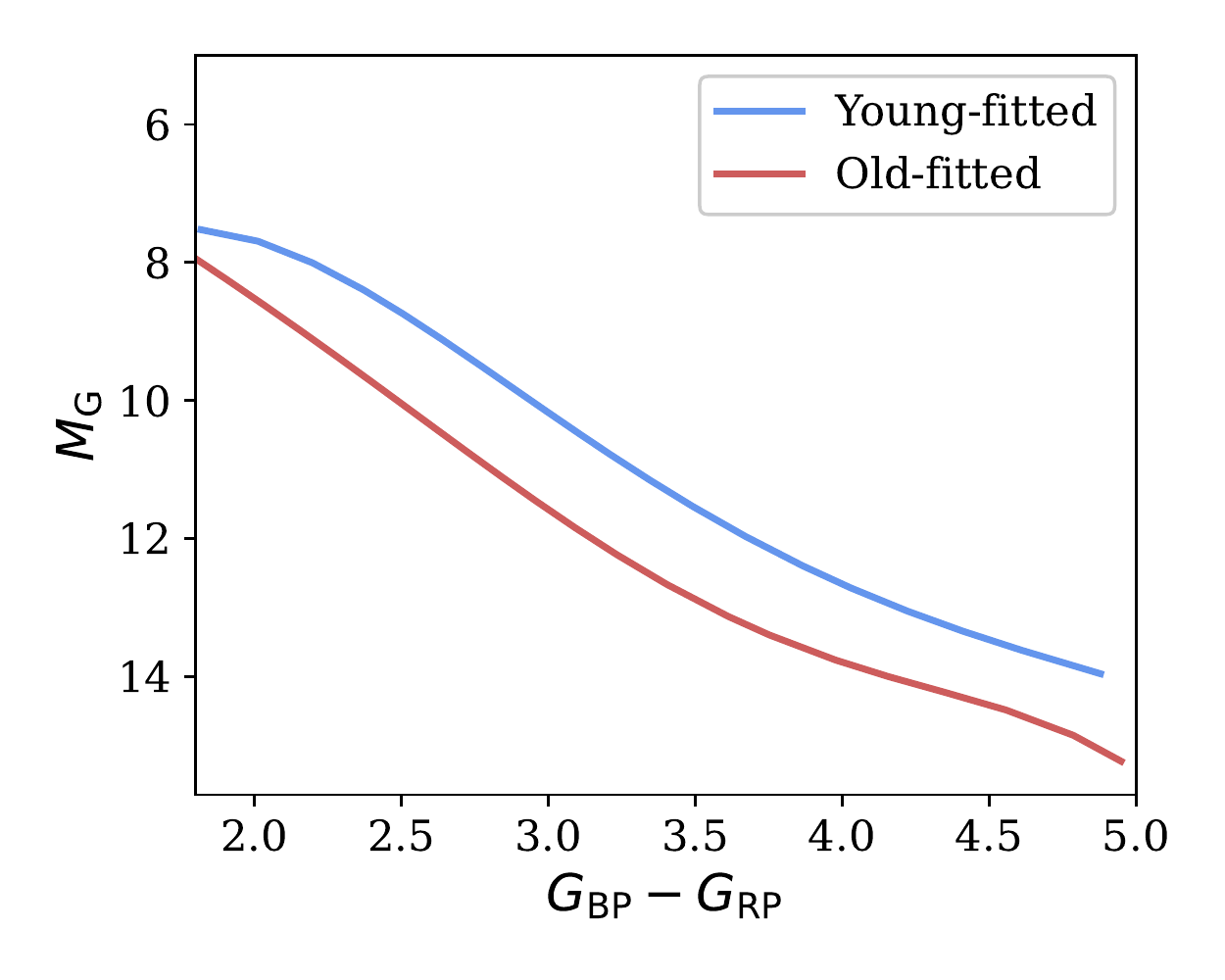}
\caption{The young and old sequences obtained from the smoothed maps in Fig.~\ref{fig:ymap2}.
These lines are 5th-order polynomial functions (equations ~\ref{eq:ymap_young} and ~\ref{eq:ymap_old}).}
\label{fig:ymap3}
\end{figure*}

In the assessment of young cool dwarfs, we generated a model map for a youth parameter [\ycmd] on a colour-magnitude diagram using three {\it Gaia} photometric bands (\gbp, \grp, and $G$).
The map is empirically derived using bona fide NYMG members and 1,000 randomly selected old stars from {\it Gaia} EDR3 (Fig.~\ref{fig:ymap1}).
In creating the \ycmd\ map, we began with two raw density maps of young and old stars.
The model maps for young and old stars were made after taking features of these two raw density maps.
First, the raw maps were smoothed by applying a Gaussian filter for eliminating bumpy features on the maps.
As a standard deviation ($\sigma$) for the Gaussian kernel, 2.0 mag was chosen empirically after testing several $\sigma$ values (e.g., 1.0 mag does not sufficiently smooth the maps, and 3.0 mag smoothes the maps too much resulting in substantial overlaps between young and old maps).
The left panel in Fig.~\ref{fig:ymap2} illustrates the smoothed maps after applying the Gaussian kernel to the raw maps. Then, the sequences of the density peak as a function of \bprp\ were found by slicing the maps.
As an example, a slice at \bprp=2.5 is illustrated in the right panel of Fig.~\ref{fig:ymap2}, displaying the \mg\  populations of young/old stars as a function of \bprp.
These sequences are well-fitted with 5th-order polynomial functions (See Fig.~\ref{fig:ymap3}).

\begin{equation}
\begin{aligned}
\!\begin{multlined}[t][.75\columnwidth]
M_{\rm G} =-0.0125 (G_{\rm BP}-G_{\rm RP})^5+0.3484 (G_{\rm BP}-G_{\rm RP})^4 \\ -3.4844 (G_{\rm BP}-G_{\rm RP})^3+15.7565 (G_{\rm BP}-G_{\rm RP})^2 \\ -30.0872 (G_{\rm BP}-G_{\rm RP})+27.5300 
  \label{eq:ymap_young}
\end{multlined}
\end{aligned}
\end{equation}

\begin{equation} 
\begin{aligned}
 \!\begin{multlined}[t][.75\columnwidth]
M_{\rm G}=0.0603 (G_{\rm BP}-G_{\rm RP})^5-0.8605 (G_{\rm BP}-G_{\rm RP})^4 \\ +4.5472 (G_{\rm BP}-G_{\rm RP})^3-11.2019 (G_{\rm BP}-G_{\rm RP})^2 \\ +15.8853 (G_{\rm BP}-G_{\rm RP})-2.9800
  \label{eq:ymap_old}
  \end{multlined}
  \end{aligned}
\end{equation}

Equations~\ref{eq:ymap_young} and ~\ref{eq:ymap_old} are the sequences of the density peaks for young and old stars, respectively.

Then, the widths of the model sequences were determined to make model maps.
As can be seen in Fig.~\ref{fig:ymap2} (right), a standard deviation of the smoothed map can be obtained at each color. Standard deviations were obtained for all color range (1.8$\leq$\bprp$<$5.0) and a mean value was calculated. Using peak sequences (equations~\ref{eq:ymap_young} and ~\ref{eq:ymap_old}) and mean values of standard deviations for young and old models  (0.607787 and  0.534968), we created model maps (Fig.~\ref{fig:ymap4}). 
After normalising and combining these two model maps, we created a \ycmd\ map having a range from 0 (old) to 1 (young), where 
a star's \ycmd\ can be obtained based on a location on the map.

\section{New candidate NYMG members}

\begin{table*}
\centering
\begin{threeparttable}
\scriptsize
\setlength\tabcolsep{3pt} 
\begin{tabular}{*{14}{c}}
\hline
 {\it Gaia} source\_id & Group & \ra\ & \dec\  &  \pmra\ & \pmdec\ & \plx   & $G$ & \bprp & $\sum{P_{\rm kin}}$& $P_{\rm kin}$ & \ycmd\ & $NUV$ & Youth(NUV)\tnote{a} \\
& & hh:mm:ss & dd:mm:ss & mas yr$^{-1}$ & mas yr$^{-1}$ & mas  & mag & mag & \% & \% & & mag &  \\ \hline    
 4683757944772826624 &      ABDor &  00:06:05.23 &  -77:56:08.0 &  125.95 &    8.09 &   19.47 & 13.35 &  2.84 &  97 &  97 & 0.96 & 22.08 & 1  \\ 
 2335390189483121152 &      ABDor &  00:13:52.41 &  -25:33:30.0 &  140.04 &   -4.83 &   23.58 & 13.04 &  2.93 &  98 &  98 & 0.98 & 20.28 & 1  \\ 
 2431769083806305152 &      ABDor &  00:19:32.06 &  -05:54:41.7 &  107.68 &  -68.02 &   22.34 & 14.65 &  3.53 &  99 &  99 & 0.98 & 22.77 & 1  \\ 
 2855743172658367360 &      ABDor &  00:24:03.96 &  +26:26:28.9 &  141.78 &  -54.49 &   26.04 & 13.29 &  2.95 &  90 &  90 & 0.88 & 23.54 & 0  \\ 
  380142611863228928 &      ABDor &  00:26:02.88 &  +39:47:23.9 &  219.14 &   21.89 &   25.61 & 14.25 &  3.21 &  92 &  92 & 0.75 & 22.14 & 1  \\ 
\multicolumn{14}{c}{\dots} \\
\multicolumn{14}{c}{\dots} \\
\multicolumn{14}{c}{\dots} \\
\hline
\end{tabular}
\begin{tablenotes}
\item[a] 0: old, 1: young, 2: indecisive of youth; the youth was evaluated using $M_{\rm NUV}$ (see Section 2.2.1 for details).
\end{tablenotes}
\end{threeparttable}
\caption{Newly suggested candidate NYMG members without RV measurements in {\it Gaia} EDR3. The entire list is available online.}
\label{tab:newcandi1}
\end{table*}


\bsp	
\label{lastpage}
\end{document}